\documentclass[a4paper,12pt]{article}
\usepackage{graphicx}
\usepackage{hyperref}
\usepackage{latexsym}
\usepackage{color}
\usepackage{amsmath}
\usepackage{multirow}
\usepackage{multicol}
\usepackage{adjustbox}
\usepackage{booktabs}
\usepackage{fancyhdr}
\usepackage{geometry}
\def\Tr{{\rm Tr}}
\def\msbar{{\overline{\rm MS}}}

\begin{document}

\title{RI/MOM and RI/SMOM renormalization of overlap quark bilinears on domain wall fermion configurations}
\author{Yujiang Bi$^1$\thanks{biyujiang@ihep.ac.cn}, Hao Cai$^1$\thanks{hcai@whu.edu.cn}, Ying Chen$^{2,3}$, \\
Ming Gong$^{2,3}$, Keh-Fei Liu$^4$, Zhaofeng Liu$^{2,3}$\thanks{liuzf@ihep.ac.cn} and Yi-Bo Yang$^{4,5}$\\
($\chi$QCD Collaboration)}

\date{}
\maketitle

\thispagestyle{fancy}
\markright{MSUHEP-17-019}

\begin{center}
$^1$School of Physics and Technology, Wuhan University, Wuhan 430072, China\\
$^2$Institute of High Energy Physics and Theoretical Physics Center for Science Facilities, Chinese Academy of Sciences, Beijing 100049, China\\
$^3$School of Physics, University of Chinese Academy of Sciences, Beijing 100049, China\\
$^4$Department of Physics and Astronomy, University of Kentucky, Lexington, KY 40506, USA\\
$^5$Department of Physics and Astronomy, Michigan State University, East Lansing, MI 48824, USA
\end{center}

\begin{abstract}
Renormalization constants (RCs) of overlap quark bilinear operators on 2+1-flavor domain wall fermion configurations are
calculated by using the RI/MOM and RI/SMOM schemes.
The scale independent RC for the axial vector current is computed by using a Ward identity. 
Then the RCs for the quark field and the vector, tensor, scalar and pseudoscalar operators are 
calculated in both the RI/MOM and RI/SMOM schemes. 
The RCs are converted to the $\msbar$ scheme and we compare the numerical results from using the two intermediate schemes.
The lattice size is $48^3\times96$ and the inverse spacing $1/a = 1.730(4) {\rm~GeV}$.
\end{abstract}

\newpage
\section{Introduction}
With the setup of overlap valence on domain wall fermion (DWF) configurations, the $\chi$QCD collaboration has
been determining the strangeness and charmness in the nucleon~\cite{Gong:2013vja}, 
the charm and strange quark masses~\cite{Yang:2014sea}, and other physical quantities of interests.
These works are based on RBC-UKQCD DWF configurations with lattice sizes $24^3\times64$ and $32^3\times64$~\cite{Aoki:2010dy,Allton:2008pn}.
To shrink uncertainties
from chiral extrapolations in calculations at unphysical light quark masses, the RBC-UKQCD Collaborations have
generated configurations at the physical pion mass on $48^3\times96$ lattices~\cite{Blum:2014tka}.
On this gauge ensemble labeled as 48I, the $\chi$QCD collaboration is studying the $\rho$ resonance~\cite{Chen:2015tpa},
nucleon magnetic moment~\cite{Sufian:2016pex,Sufian:2017osl},
and decay constants of pseudoscalar and vector mesons~\cite{Weifeng}.
To link hadronic matrix elements computed on the lattice to the continuum world, we need the RCs for the corresponding operators.
In this paper we present our calculation of the RCs for the flavor nonsinglet scalar ($S$), pseudoscalar ($P$), vector ($V$), 
axial vector ($A$) and tensor ($T$) currents of
overlap valence quark on the 48I ensemble. The quark field RC is also obtained.

The RI/MOM scheme~\cite{Martinelli:1994ty} is a popular nonperturbative method to calculate RCs in lattice calculations.
The results are then converted to the $\msbar$ scheme by using conversion ratios from perturbation theory.
With the shrink of statistical uncertainties in RCs of flavor nonsinglet quark bilinears, 
the truncation error in the conversion ratio from the RI/MOM scheme to the
$\msbar$ scheme starts to dominate the total uncertainty of the RC $Z_S(=1/Z_m)$. To reduce this truncation error, 
the RI/SMOM scheme~\cite{Aoki:2007xm,Sturm:2009kb} was proposed in which unexceptional or symmetric momentum modes are used
when calculating vertex functions of operators.
The conversion ratio from the RI/SMOM scheme to the $\msbar$ 
scheme for the scalar density was shown to converge much faster than in the case of RI/MOM scheme~\cite{Gorbahn:2010bf,Almeida:2010ns}.
Also, the nonperturbative effects from chiral symmetry breaking and other infrared effects are expected to be more suppressed 
in the RI/SMOM scheme~\cite{Aoki:2007xm}.

In this work we compute the aforementioned RCs by using both the RI/MOM and RI/SMOM schemes. 
In the end the RCs are converted to the $\msbar$ scheme.
The numerical results are compared to try to see the
advantages and shortcomings of the two intermediate schemes.
After converting to the $\msbar$ scheme, we perform perturbative runnings and 
give the results at 2 GeV for the scale dependent RCs $Z_q$, $Z_S$, $Z_P$ and $Z_T$.
Throughout this paper we use the conventions below for the RCs of the quark field, quark mass and bilinear operators:
\begin{equation}
\psi_R=Z_q^{1/2}\psi_R,\quad m_R=Z_m m_B,\quad \mathcal{O}_R=Z_{\mathcal{O}}\mathcal{O}_B,
\end{equation}
where the subscripts $R$ and $B$ denote the renormalized and bare quantities respectively.

This paper is organized as follows. In Sec.~\ref{sec:framework} we give our framework of the calculation, including the definitions
of the renormalization schemes, our overlap fermion Dirac operator and the information of the gauge configurations. Sec.~\ref{sec:details} shows
the computation details, the numerical results and discussions. Finally we summarize in Sec.~\ref{sec:summary}.

\section{Framework of our calculation}
\label{sec:framework}
In both the RI/MOM and RI/SMOM schemes, the renormalization condition for an operator is imposed on its amputated Green function
in the vanishing quark mass limit.
The Green function $G_{\mathcal{O}}$ is computed between two external off-shell quark states in Landau gauge. If using a point source
quark propagator, one has
\begin{align}
  \begin{split}
	G_{\mathcal{O}}(p_1,p_2)=&\sum_{x,y}e^{-i(p_1\cdot x-p_2\cdot y)}\left<\psi(x)\mathcal{O}{(0)\bar{\psi}(y)}\right>,
  \end{split}
\end{align}
where $\mathcal{O}=\bar\psi\Gamma\psi$ with 
$\Gamma=I, \gamma_5, \gamma_\mu, \gamma_\mu\gamma_5, \sigma_{\mu\nu}(=\frac{1}{2}[\gamma_\mu,\gamma_\nu])$.
The amputated Green function is then
\begin{equation}
  \Lambda_{\mathcal{O}}(p_1,p_2)=S^{-1}(p_1)G_{\mathcal{O}}(p_1,p_2)S^{-1}(p_2),
\end{equation}
where the quark propagator $S(p)$ in momentum space is
\begin{equation}
  S(p)=\sum_{x} e^{-ip\cdot x}\langle \psi(x)\bar{\psi}(0)\rangle.
\end{equation}

In the RI/MOM scheme, one uses the forward Green function. That is to say, the momenta satisfy $p_1=p_2=p$. The renormalization
condition is imposed at the scale $p_1^2=p_2^2=p^2=\mu^2$ by
\begin{equation}
\label{eq:mom}
\underset{m_R\to 0}{\text{lim}}
Z^{-1}_qZ_{\mathcal{O}}\frac{1}{12}\Tr[\Lambda_{\mathcal{O},B}(p)\Lambda_{\mathcal{O}}^{\text{tree}}(p)^{-1}]_{p^2=\mu^2}=1,
\end{equation}
where the subscript $B$ stands for bare and the projector $\Lambda_{\mathcal{O}}^{\rm tree}(p)=\Gamma$ for the quark bilinears considered in this work. 
The quark field RC in the RI/MOM scheme is determined by
\begin{equation}
Z_q^{\rm RI/MOM}(\mu)=\underset{m_R\to 0}{\text{lim}}\frac{-i}{48}\Tr\left[\gamma_\nu\frac{\partial S^{-1}(p)}{\partial p_\nu}\right]_{p^2=\mu^2},
\end{equation}
which is compatible with the vector Ward-Takahashi identity. To avoid the inconvenience caused by the derivative with
respect to the discretized momenta on the lattice, we use the RC for the local axial vector current $Z_A^{\rm RI/MOM}$ as the input to calculate
other RCs. For example, from Eq.(\ref{eq:mom}) the quark field RC can be obtained by
\begin{equation}
Z_q^{\rm RI/MOM}=Z_A^{\rm RI/MOM}\underset{m_R\to 0}{\text{lim}}
\frac{1}{12}\Tr\left[\Lambda_{A,B}(p)\Lambda_{A}^{\rm tree}(p)^{-1}\right]_{p^2=\mu^2}.
\label{eq:zq_mom}
\end{equation}
At large $\mu$ the renormalization condition for $Z_A^{\rm RI/MOM}$ is compatible with the axial vector 
Ward-Takahashi identity~\cite{Martinelli:1994ty}. Thus $Z_A^{\rm RI/MOM}$ equals to a value $Z_A^{\rm WI}$ 
obtained from some Ward identity on hadron states.
In our work below, we shall use the partially conserved axial current (PCAC) relation to determine $Z_A^{\rm WI}$.
We will also use the relation $Z_A^{\rm WI}=Z_A^\msbar$ in this work since the $\msbar$ scheme is consistent with the chiral
Ward identities too.

In the RI/SMOM scheme~\cite{Sturm:2009kb}, one uses the symmetric momentum configuration
\begin{equation}
q^2\equiv(p_1-p_2)^2=p_1^2=p_2^2=\mu^2
\label{eq:sym_mom}
\end{equation}
when fixing the RCs at the scale $\mu$. The projectors for the amputated Green functions of the scalar, pseudoscalar and tensor currents
are the same as those in the RI/MOM scheme. But for the vector and axial vector currents the conditions are~\cite{Sturm:2009kb}
\begin{eqnarray}
  \underset{m_R\to 0}{\text{lim}}Z_q^{-1}Z_V\frac{1}{12q^2}\text{Tr}[q_{\mu}\Lambda_{V,B}^{\mu}(p_1,p_2)q\!\!\!/]_{\rm sym}&=1,
  \label{eq:zv_smom}\\
  \underset{m_R\to 0}{\text{lim}}Z_q^{-1}Z_A\frac{1}{12q^2}\text{Tr}[q_{\mu}\Lambda_{A,B}^{\mu}(p_1,p_2)\gamma_5q\!\!\!/]_{\rm sym}&=1.
  \label{eq:za_smom}
\end{eqnarray}
Here the subscript ``sym" denotes the symmetric momentum configuration in Eq.(\ref{eq:sym_mom}).
The quark field RC in the RI/SMOM scheme is given by
\begin{equation}
  Z_q^{\rm RI/SMOM}=\underset{m_R\to 0}{\text{lim}}\frac{1}{12p^2}\text{Tr}[S_B^{-1}(p)p\!\!\!/]_{p^2=\mu^2},
  \label{eq:zq_smom}
\end{equation}
which is the same as that in the RI'/MOM scheme~\cite{Martinelli:1994ty}. The conditions in 
Eqs.(\ref{eq:zv_smom},\ref{eq:za_smom},\ref{eq:zq_smom})
are compatible with the vector and axial vector Ward-Takahashi identities~\cite{Sturm:2009kb}. Therefore one has 
$Z_A^{\rm RI/SMOM}=Z_A^{\rm WI}$. Then by using Eq.(\ref{eq:za_smom}), we can, alternatively, obtain
\begin{equation}
 Z_q^{\rm RI/SMOM}=\underset{m_R\to 0}{\text{lim}}Z_A^{\rm WI}
 \frac{1}{12q^2}\text{Tr}[q_{\mu}\Lambda_{A,B}^{\mu}(p_1,p_2)\gamma_5q\!\!\!/]_{\rm sym}.
 \label{eq:zq_smom_wi}
\end{equation}

In using the RI/MOM and RI/SMOM schemes in practical lattice calculations at a lattice spacing $a$, one needs a renormalization window
\begin{equation}
 \Lambda_{\rm QCD}\ll \mu \ll \pi/a,
\end{equation}
in which both the infrared effects from chiral symmetry breaking and the ultraviolet effects from the lattice cutoff are small.
Also perturbation theory can only apply at large enough momentum scale for 
calculating the conversion ratios of RCs to the $\msbar$ scheme.

We use overlap fermions~\cite{Neuberger:1997fp} as the valence quark. Our massless overlap operator is given by
\begin{equation}
D_{ov}  (\rho) =   1 + \gamma_5 \varepsilon (\gamma_5 D_{\rm w}(\rho)),
\end{equation}
where $\varepsilon$ is the matrix sign function and $D_{\rm w}(\rho)$ is the usual Wilson fermion operator, 
except with a negative mass parameter $- \rho = 1/2\kappa -4$ in which $\kappa_c < \kappa < 0.25$. 
$\kappa$ is set to $0.2$ in our calculation, which corresponds to $\rho = 1.5$. The massive overlap Dirac operator is defined as
\begin{eqnarray}
D_m &=& \rho D_{ov} (\rho) + m\, (1 - \frac{D_{ov} (\rho)}{2}) \nonumber\\
    &=& \rho + \frac{m}{2} + (\rho - \frac{m}{2})\, \gamma_5\, \varepsilon (\gamma_5 D_w(\rho)).
\end{eqnarray}
To accommodate the SU(3) chiral transformation, we use the chirally regulated field
$\hat{\psi} = (1 - \frac{1}{2} D_{ov}) \psi$ in place of $\psi$ in the interpolation field and the currents.
This amounts to leave the unmodified currents and instead adopt the effective propagator
\begin{equation}
G \equiv D_{eff}^{-1} \equiv (1 - \frac{D_{ov}}{2}) D^{-1}_m = \frac{1}{D_c + m},
\end{equation}
where $D_c = \frac{\rho D_{ov}}{1 - D_{ov}/2}$ is chiral, i.e. $\{\gamma_5, D_c\}=0$~\cite{Chiu:1998gp}.
With the good chiral properties of overlap fermions,
we should expect $Z_S=Z_P$ and $Z_V=Z_A$. These relations are indeed satisfied within uncertainties 
by our numerical results as will be shown later. We also expect that the RI/SMOM results satisfy these relations
better than the RI/MOM results since the RI/SMOM scheme suppresses more nonperturbative effects
from chiral symmetry breaking.

The gauge configurations that we use in this work are from the RBC-UKQCD Collaborations~\cite{Blum:2014tka}.
2+1 flavor domain wall fermions were used as the sea quarks in generating these configurations.
The light sea quark mass is essentially at the physical point.
The lattice size is $48^3\times96$ and the inverse lattice spacing is $1/a = 1.730(4) {\rm~GeV}$. This ensemble
is called 48I by the RBC-UKQCD Collaborations. The parameters of these configurations are collected in Tab.~\ref{tab:nconfs}.
\begin{table}
\begin{center}
\caption{Parameters of configurations with 2+1 flavor dynamical domain wall fermions (RBC-UKQCD Collaborations~\cite{Blum:2014tka}).
The number of configurations used in this work is 81. The residual mass is read from Fig.~6 in~\cite{Blum:2014tka}.}
\begin{tabular}{cccccc}
\hline\hline
$1/a$(GeV) & label & $am_l^{sea}/am_s^{sea}$ & volume & $N_{\rm conf}$ & $am_{\rm res}$ \\
\hline
1.730(4) & 48I  & 0.00078/0.0362 & $48^3\times96$ & $81$ & 0.000610(4)\\
\hline\hline
\end{tabular}
\label{tab:nconfs}
\end{center}
\end{table}

In Tab.~\ref{tab:valence} we give the overlap valence quark masses in lattice units used in this work.
\begin{table}
\begin{center}
\caption{Overlap valence quark masses in lattice units used in this work. The corresponding pion masses are from Ref.~\cite{Chen:2015tpa}.}
\begin{tabular}{ccccccccc}
\hline\hline
$am_q$  & 0.00120 & 0.00170 & 0.00240 & 0.00300 & 0.00455 & 0.00600 & 0.0102 & 0.0203 \\
\hline
$m_\pi$/MeV &  95(3)  & 114(2) & 135(2) & 149(2)  &  182(2) & 208(2) & 267(1) & 371(1) \\ 
\hline\hline
\end{tabular}
\label{tab:valence}
\end{center}
\end{table}
The corresponding pion masses in the table were measured in Ref.~\cite{Chen:2015tpa} and are close to the chiral limit. 
The precise values of the pion masses do not matter here
since we will use the valence quark mass to do the extrapolations to the chiral limit.

\section{Calculation and numerical results}
\label{sec:details}
We use periodic boundary conditions in all four directions. Therefore the discretized momenta in lattice units are
\begin{equation}
ap=2\pi\left(\frac{k_1}{L}, \frac{k_2}{L}, \frac{k_3}{L}, \frac{k_4}{T}\right),
\end{equation}
where $L=48$, $T=96$ and $k_\mu$ are integers. In doing the Fourier transformation for the point source quark propagators, 
we set $k_\mu=-12,-11,...,12$. To reduce the effects of Lorentz
non-invariant discretization errors, we only use the momenta which satisfy the ``democratic" condition
\begin{equation}
\frac{p^{[4]}}{(p^2)^2}<0.29,\quad\mbox{where } p^{[4]}=\sum_\mu p_\mu^4,\quad p^2=\sum_\mu p_\mu^2
\label{eq:p_cut}
\end{equation}
in performing the RI/MOM scheme calculation. In other words, only those momenta aligning along or close to the 
4-dimensional diagonal line are used for the RI/MOM analyses. 
For the RI/SMOM calculation, the conditions in
Eq.(\ref{eq:sym_mom}) cannot be easily satisfied. Thus we do not apply any ``democratic" cuts like the one in Eq.(\ref{eq:p_cut}) for this case.
We use point source quark propagators in Landau gauge to compute all the necessary gauge dependent Green functions and vertex functions.
The statistical errors of our numerical results are from Jackknife processes with one configuration removed each time.

After obtaining the RCs in the RI/MOM and RI/SMOM schemes at each scale $\mu$, we convert them to their $\msbar$ values by using the
corresponding conversion ratios calculated in perturbation theory at that scale. Then a perturbative running to 2 GeV 
in the $\msbar$ scheme is performed for each RC by using the appropriate anomalous dimensions.

\subsection{Renormalization of the axial vector current from PCAC}
\label{sec:zawi}
Similar to what was done in Ref.~\cite{Liu:2013yxz}, we use the PCAC relation
\begin{equation}
Z_A\partial_\mu A_\mu=2Z_m m_q Z_P P,
\label{eq:ZA_WI}
\end{equation}
and $Z_m=Z_P^{-1}$ for overlap fermions to obtain $Z_A^{\rm WI}$.
By sandwiching both sides of Eq.(\ref{eq:ZA_WI}) into the vacuum and a pion state at rest, one finds
\begin{equation}
Z_A^{\rm WI}=\frac{2m_q\langle\Omega|P|\pi\rangle}{m_\pi\langle \Omega| A_4 |\pi\rangle}.
\end{equation}
To get the ratio of matrix elements and the pion mass, 
we calculate zero momentum 2-point correlators $\langle\Omega|P P^\dagger|\Omega\rangle$ and
$\langle\Omega|A_4 P^\dagger|\Omega\rangle$ in practice by using wall-source quark propagators.
For a given $m_q$, we simultaneously fit the two wall-source point-sink correlators at large source-sink time separation by a single exponential
with $m_\pi$ as a common parameter. The ratio of the matrix elements is then given by the ratio of the spectral weights
(the other two fitting parameters) in front of
the exponentials.

\begin{figure}[htbp]
\centering \includegraphics[]{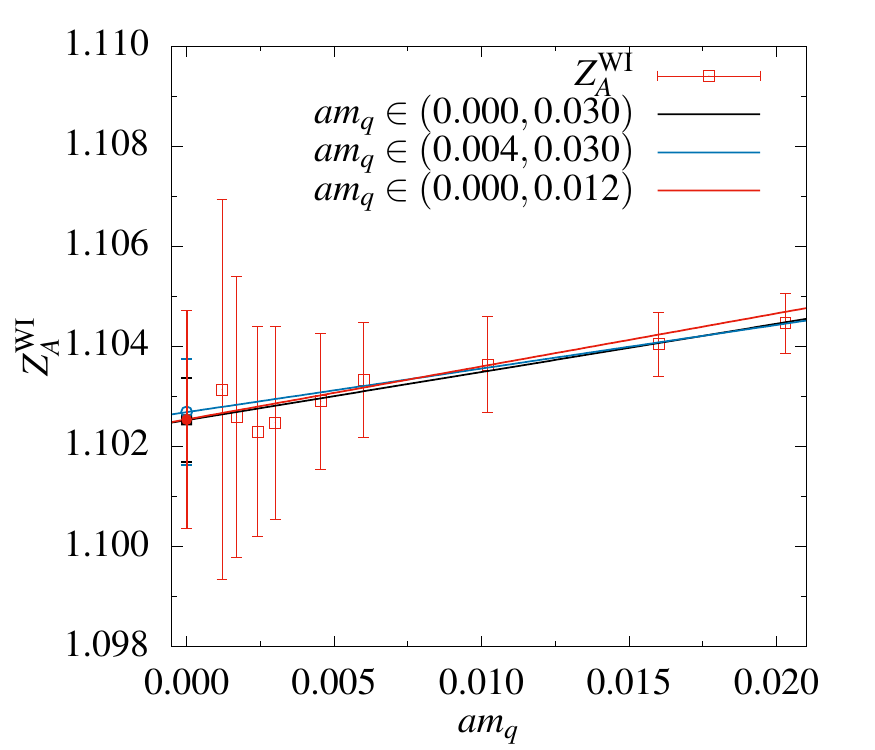}
 \caption{$Z_A^{\rm WI}$ as a function of the valence quark mass and its linear chiral extrapolations using three different
 fitting ranges.}
 \label{fig:za_wi}
\end{figure}

The resulted $Z_A^{\rm WI}$ is plotted as a function of $am_q$ in Fig.~\ref{fig:za_wi}, 
in which we also plot
the linear extrapolation of $Z_A^{\rm WI}$ to the chiral limit.
The numerical values of $Z_A^{\rm WI}$ are given in Tab.~\ref{tab:zawi}.
\begin{table}[htbp]
\begin{center}
\caption{$Z_A^{\text{WI}}$ for 9 valence quark masses. $Z_A^{\rm WI}$ at $am_q=0$ is from a linear extrapolation in $am_q$.
The first error is statistical and the second is a systematic error from varying the range of $am_q$ in doing the linear
chiral extrapolation.}
\begin{tabular}{cccccc}
\hline\hline
$am_q$ & 0.0 & \hspace{0.3em}0.00120 & \hspace{0.3em}  0.00170 &  \hspace{0.3em} 0.00240 &  \hspace{0.3em}0.00300  \\
$Z_A^{\rm WI}$ & 1.1025(8)(4)    & 1.1031(38) & 1.1026(28) & 1.1023(21) & 1.1025(19)  \\
\hline
$am_q$  &  \hspace{0.3em}0.00455 &  \hspace{0.3em}0.00600 & \hspace{0.3em} 0.01020 & \hspace{0.3em} 0.01600 & \hspace{0.3em} 0.02030  \\
$Z_A^{\rm WI}$ &  1.1029(14) & 1.1033(12) & 1.1036(10) & 1.1041(6) & 1.1045(6)\\
\hline\hline
\end{tabular}
\label{tab:zawi}
\end{center}
\end{table}
By using all the 9 data points at nonzero $am_q$ in Tab.~\ref{tab:zawi} we obtain $Z_A^{\rm WI}=1.1025(8)$
in the chiral limit, where the error is
only statistical.
The systematic error of $Z_A^{\rm WI}$ in the chiral limit is determined by varying the range of $am_q$ in doing the chiral extrapolation.
Using the data points at $am_q>0.004$ we get $Z_A^{\rm WI}=1.1027(11)$. If we drop the data at the largest two quark masses
$am_q=0.00160$ and $0.02030$, then we find $Z_A^{\rm WI}=1.1023(13)$. 
In the end we assign the largest difference $0.0004$ in these center values
as the systematic error. Combining the two errors quadratically, we get $Z_A^{\rm WI}=1.1025(9)$.

\subsection{The vector current}
\begin{figure}[htpb]
\centering\includegraphics[width=\textwidth]{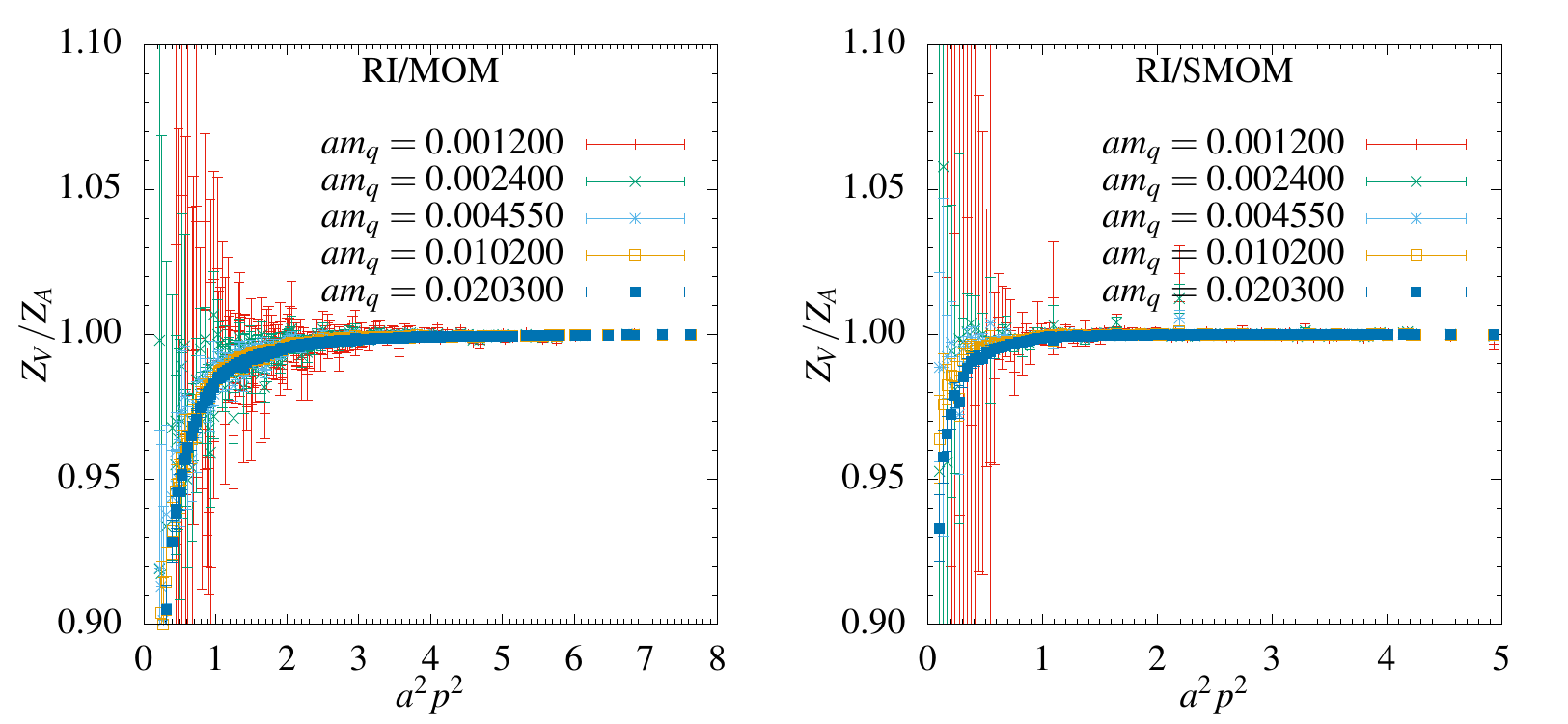}
\caption{$Z_V/Z_A$ in the RI/MOM (left graph) and RI/SMOM (right graph) scheme. The results show little valence quark mass dependence.}
\label{fig:zv_massive}
\end{figure}

\begin{figure}[htpb]
\centering\includegraphics[width=0.5\textwidth]{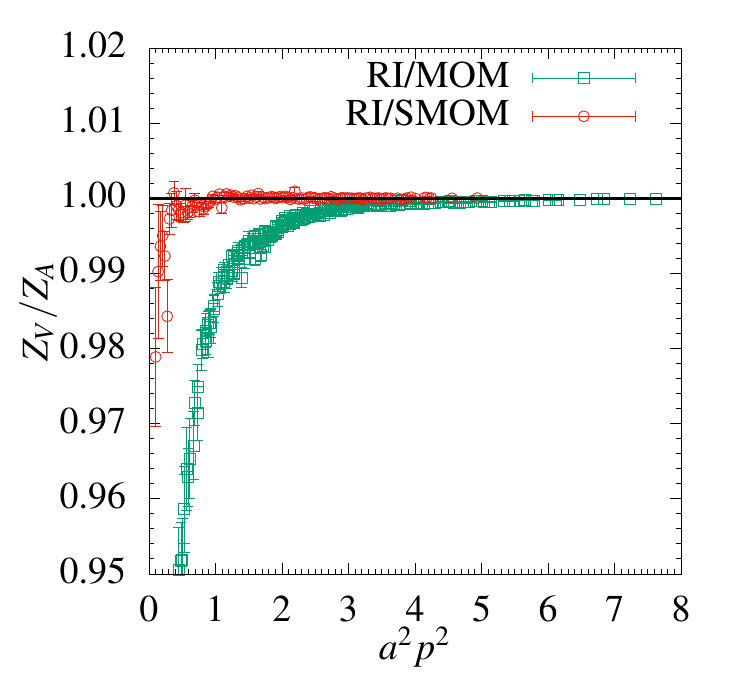}
\caption{$Z_V/Z_A$ in the chiral limit in the two schemes. $Z_V/Z_A=1$ is well satisfied in both schemes at large momentum scale.
This relation is verified at lower momentum scale in the RI/SMOM scheme than in the RI/MOM scheme. 
The horizontal line is $Z_V/Z_A=1$ for guiding the eyes.}
\label{fig:zv_massless}
\end{figure}

For overlap fermions, $Z_V=Z_A$ is expected from its good chiral property. We calculate the ratio $Z_V/Z_A$ in both
RI/MOM and RI/SMOM schemes:
\begin{equation}
\frac{Z_V^{\rm RI/MOM}}{Z_{A}^{\rm RI/MOM}}=\left.\frac{\Gamma_A(p)}{\Gamma_V(p)}\right|_{p^2=\mu^2},\quad
\frac{Z_V^{\rm RI/SMOM}}{Z_{A}^{\rm RI/SMOM}}=\left.\frac{\Gamma_A(p_1,p_2)}{\Gamma_V(p_1,p_2)}\right|_{\rm sym},
\end{equation}
where
\begin{equation}
\Gamma_V(p)=\frac{1}{48}\Tr[\Lambda_{V,B}^{\mu}(p)\gamma_\mu],
\quad \Gamma_A(p)=\frac{1}{48}\Tr[\Lambda_{A,B}^{\mu}(p)\gamma_5\gamma_\mu],
\end{equation}
\begin{equation}
\Gamma_V(p_1,p_2)=\frac{1}{12q^2}\Tr[q_{\mu}\Lambda_{V,B}^{\mu}(p_1,p_2)q\!\!\!/], \quad
\Gamma_A(p_1,p_2)=\frac{1}{12q^2}\Tr[q_{\mu}\Lambda_{A,B}^{\mu}(p_1,p_2)\gamma_5q\!\!\!/].
\end{equation}
The numerical results of this ratio are shown in Fig.~\ref{fig:zv_massive} for some of the valence quark masses $am_q$. 
Clearly little quark mass
dependence is seen in these results.
We do linear extrapolations in $am_q$ for $Z_V/Z_A$ in both schemes to reach the chiral limit.
The comparison of this ratio in the two schemes in the chiral limit is shown in Fig.~\ref{fig:zv_massless}.
$Z_V/Z_A=1$ is well satisfied in both schemes at large momentum scale. The RI/SMOM scheme is supposed to have less infrared effects.
Indeed, $Z_V/Z_A=1$ is verified at lower momentum scale in the RI/SMOM scheme than in the RI/MOM scheme as shown in Fig.~\ref{fig:zv_massless}.

\subsection{Quark field renormalization}\label{sec:zq}
\begin{figure*}[!htbp]
 \includegraphics[width=\textwidth]{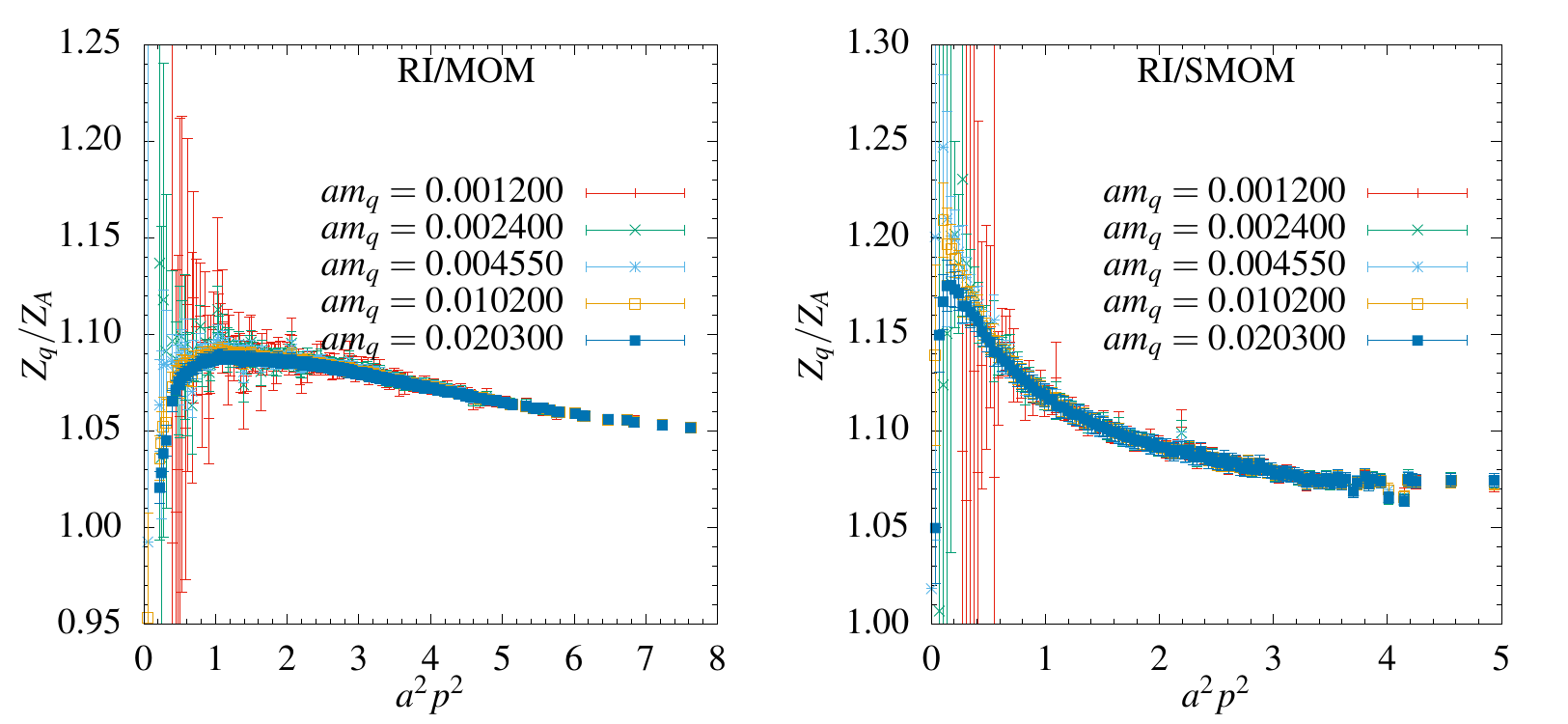}
 \caption{$Z_q/Z_A$ in the RI/MOM and RI/SMOM schemes at various valence quark masses.}
 \label{fig:zq_mom_smom}
\end{figure*}

\begin{figure}[!htpb]
  \includegraphics[width=\textwidth]{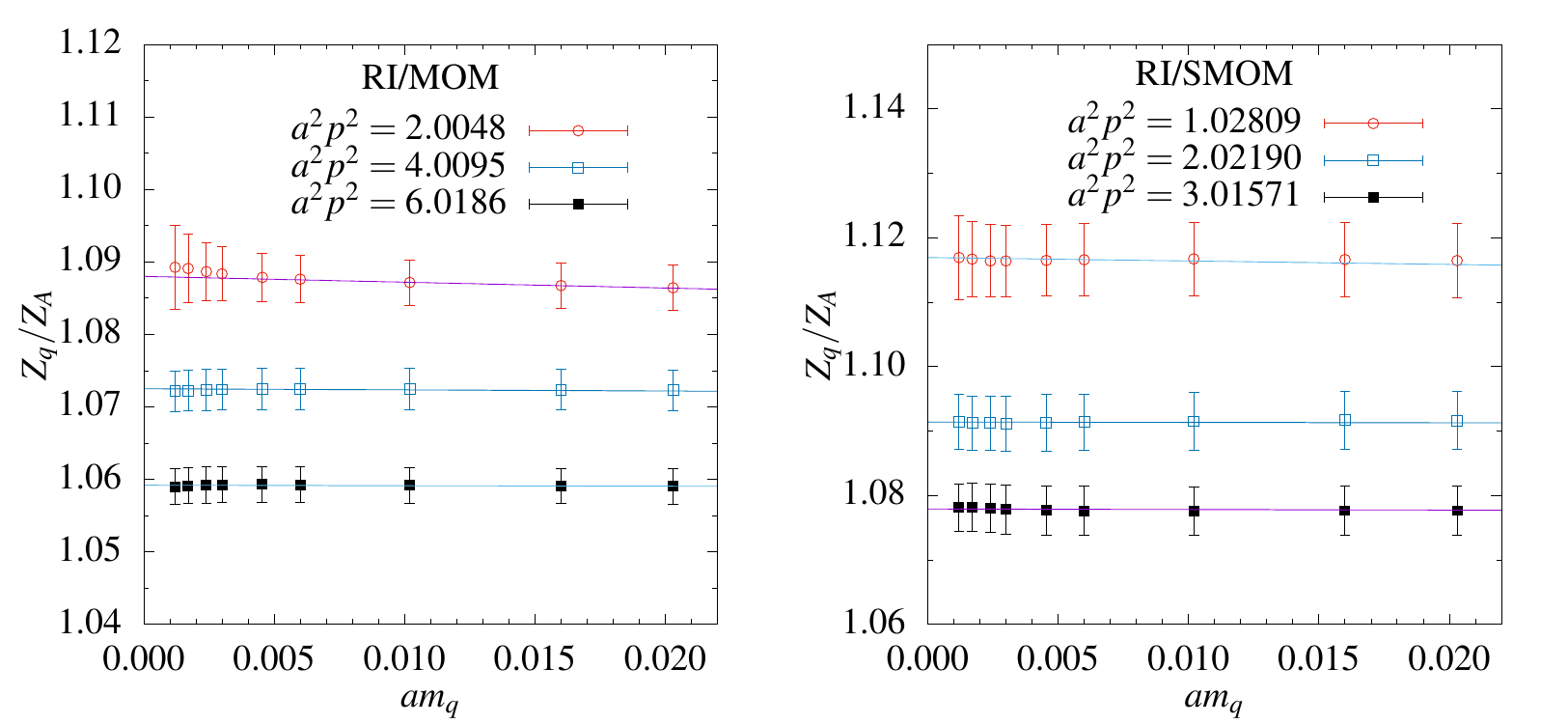}
  \caption{Linear chiral extrapolations of $Z_q/Z_A$ in the RI/MOM and RI/SMOM schemes at three typical momentum values.}
    \label{fig:zq_amq_linear}
\end{figure}

The quark field RC can be used, for example, in analyzing the scalar dressing function of the quark propagator to determine
the quark chiral condensate~\cite{Wang:2016lsv}. After finding $Z_A^{\rm WI}$, we use Eqs.(\ref{eq:zq_mom},\ref{eq:zq_smom_wi}) to calculate
$Z_q$ in the RI/MOM and RI/SMOM scheme respectively. What we get are shown in Fig.~\ref{fig:zq_mom_smom}, in which $Z_q/Z_A$ is plotted as
a function of the renormalization scale for various valence quark masses.
Apparently the quark mass dependence of $Z_q/Z_A$ in both schemes is very small. The chiral extrapolation can be done with a linear function
\begin{equation}
\frac{Z_q}{Z_A}(am_q)=\frac{Z_q}{Z_A}+A\cdot am_q.
\label{eq:zq_amq_linear}
\end{equation}
Examples of this linear extrapolation of $Z_q/Z_A$ in the RI/MOM and RI/SMOM schemes are shown in the two graphs in Fig.~\ref{fig:zq_amq_linear}.

The chiral limit value of $Z_q^{\rm RI/MOM}/Z_A$ can be converted to the $\msbar$ value $Z_q^{\msbar}/Z_A$ by 
using the following 3-loop conversion ratio 
given in Ref.~\cite{Chetyrkin:1999pq} (we use the relation $Z_A^{\rm WI}=Z_A^{\rm RI/MOM}=Z_A^{\rm RI/SMOM}=Z_A^{\msbar}$ here
and in the rest of the paper)
\begin{eqnarray}
\frac{Z_q^\msbar}{Z_q^{\text{RI/MOM}}}&=&1+\left[-\frac{517}{18}+12\zeta_3+\frac{5}{3}n_f\right]\left(\frac{\alpha_s}{4\pi}\right)^2
  +\left[-\frac{1287283}{648}+\frac{14197}{12}\zeta_3 +\frac{79}{4}\zeta_4\right.\nonumber\\
  &&\left.-\frac{1165}{3}\zeta_5
    +\frac{18014}{81}n_f-\frac{368}{9}\zeta_3n_f-\frac{1102}{243}n_f^2\right]\left(\frac{\alpha_s}{4\pi}\right)^3+\mathcal{O}(\alpha_s^4).
\label{eq:zq_mom_convert}
\end{eqnarray}
Here $n_f$ is the number of flavors and $\zeta_n$ is the Riemann 
zeta function evaluated at $n$. The strong coupling constant $\alpha_s(\mu)$ is evaluated in the $\msbar$ scheme by
using its perturbative running to four loops~\cite{Alekseev:2002zn}.
The beta functions in the $\msbar$ scheme to 4-loops can be found in Ref.~\cite{vanRitbergen:1997va}. 
And we use $\Lambda_{\rm QCD}^\msbar=332(17)$ MeV for three flavors in the $\msbar$ scheme~\cite{Olive:2016xmw}.

Since $Z_q^{\rm RI/SMOM}=Z_q^{\rm RI'/MOM}$, the conversion of $Z_q^{\rm RI/SMOM}/Z_A$ to its $\msbar$ value 
is done with the ratio~\cite{Chetyrkin:1999pq}
\begin{eqnarray}
\frac{Z_q^\msbar}{Z_q^{\text{RI'/MOM}}}&=&1+\left[-\frac{359}{9}+12\zeta_3+\frac{7}{3}n_f\right]\left(\frac{\alpha_s}{4\pi}\right)^2
  +\left[-\frac{439543}{162}+\frac{8009}{6}\zeta_3 +\frac{79}{4}\zeta_4\right.\nonumber\\
  &&\left.-\frac{1165}{3}\zeta_5+\frac{24722}{81}n_f-\frac{440}{9}\zeta_3 n_f-\frac{1570}{243}n_f^2\right]
  \left(\frac{\alpha_s}{4\pi}\right)^3+\mathcal{O}(\alpha_s^4)
\label{eq:zq_smom_convert}
\end{eqnarray}

\begin{figure}[htpb]
\centering\includegraphics[width=\textwidth]{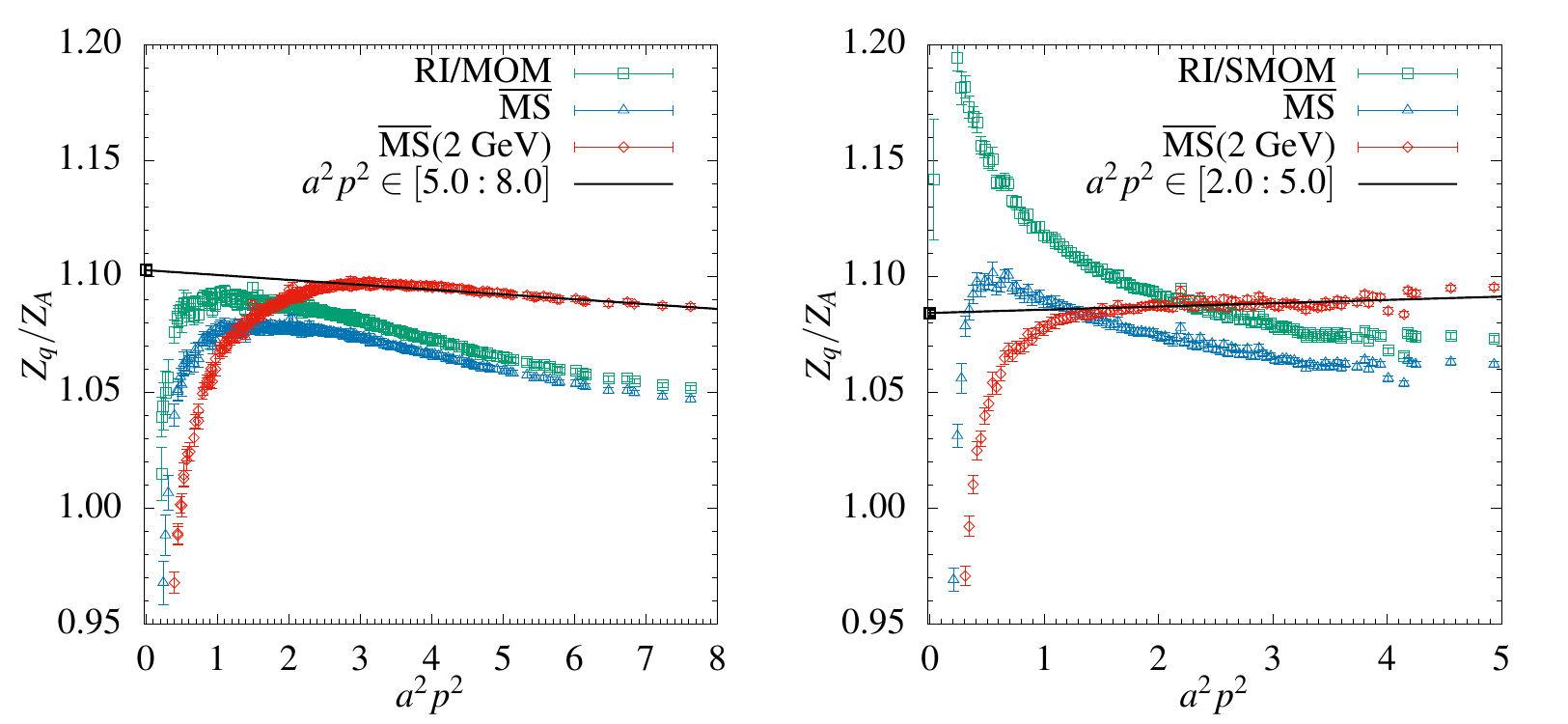}
\caption{Conversion of $Z_q/Z_A$ in the RI/MOM or RI/SMOM scheme to the $\msbar$ scheme. The running to 2 GeV in
the $\msbar$ scheme is also shown by the red diamonds in both graphs. The black line in each graph is a linear extrapolation in $a^2p^2$
using data in the indicated range.}
\label{fig:zq_za_ap2}
\end{figure}

The conversions of $Z_q^{\rm RI/MOM}/Z_A$ and $Z_q^{\rm RI/SMOM}/Z_A$ are plotted in the two graphs respectively 
in Fig.~\ref{fig:zq_za_ap2}.
The Green squares are the results in the momentum subtraction schemes. After converting them to the $\msbar$ scheme,
we obtain the blue triangles.

Note the perturbative truncation error in the conversion ratio Eq.(\ref{eq:zq_smom_convert}) for the RI/SMOM scheme
is large than that in Eq.(\ref{eq:zq_mom_convert}) for the RI/MOM scheme. 
For example, at $p^2=\mu^2=16$ GeV$^2$ (or $a^2p^2=5.346$ with our lattice spacing)
the numerical value of Eq.(\ref{eq:zq_mom_convert})
can be broken into
\begin{eqnarray}
\frac{Z_q^{\msbar}}{Z_q^{\rm RI/MOM}}(\mu=4\mbox{ GeV},n_f=3)&=&1-0.0\alpha_s-0.0589\alpha_s^2-0.2352\alpha_s^3+\cdots\nonumber\\
&=&1-0.0-0.0028-0.0025+\cdots,
\end{eqnarray}
where we have used $\alpha_s^\msbar(4\mbox{ GeV})=0.2189$. 
Assuming the coefficient of the $\mathcal{O}(\alpha_s^4)$ term is 4($\approx 0.2352/0.0589$) times larger
than that of the $\mathcal{O}(\alpha_s^3)$ term, we can estimate the $\mathcal{O}(\alpha_s^4)$ term to be of size $\sim0.0022$. This
means the truncation error is of size $0.2\%$. At the same scale, the numerical value of Eq.(\ref{eq:zq_smom_convert}) is
\begin{eqnarray}
\frac{Z_q^{\msbar}}{Z_q^{\rm RI'/MOM}}(\mu=4\mbox{ GeV},n_f=3)&=&1-0.0\alpha_s-0.1169\alpha_s^2-0.4076\alpha_s^3+\cdots\nonumber\\
&=&1-0.0-0.0056-0.0043+\cdots.
\end{eqnarray}
Assuming the coefficient of the $\mathcal{O}(\alpha_s^4)$ term is 3.5($\approx 0.4076/0.1169$) times larger
than that of the $\mathcal{O}(\alpha_s^3)$ term, we find that the size of the $\mathcal{O}(\alpha_s^4)$ term is about 0.0033.
Thus the truncation error is $0.3\%$.

Our RI/SMOM scheme data do not reach beyond the scale $a^2p^2=\sim5$ as shown in the right panel of Fig.~\ref{fig:zq_za_ap2}. 
We use the data starting from $a^2p^2=2$ (or $p=2.447$ GeV) in our analyses below.
The truncation error in Eq.(\ref{eq:zq_smom_convert}) at $a^2p^2=2$ can be estimated similarly and its size is $0.7\%$.

The $\msbar$ value $Z_q^\msbar/Z_A$ at a given scale $a^2p^2$ can be 
run to 2 GeV by using the quark field anomalous dimension $\gamma_q^\msbar$. In perturbation theory
$\gamma_q^\msbar$ has been calculated to 4-loops in Landau gauge~\cite{Chetyrkin:1997dh}.
The red diamonds in both graphs of Fig.~\ref{fig:zq_za_ap2} show $Z_q^\msbar/Z_A(2\mbox{ GeV}; a^2p^2)$ after the running
as a function of the initial scale $a^2p^2$.
From the linear dependence on $a^2p^2$ at large scale we extrapolate $Z_q^\msbar/Z_A(2\mbox{ GeV}; a^2p^2)$ to $a^2p^2=0$
to remove the $\mathcal{O}(a^2p^2)$ lattice artefacts.

In the left graph of Fig.~\ref{fig:zq_za_ap2} using the RI/MOM scheme, we do the linear extrapolation in the range
$a^2p^2\in [5,8]$ and find $Z_q^\msbar/Z_A(2\mbox{ GeV})=1.1027(20)$. If using the range $a^2p^2\in [4,8]$, 
then we get $Z_q^\msbar/Z_A(2\mbox{ GeV})=1.1052(11)$.
The change in the center value ($0.2\%$) is taken as a systematic error. 

For the right graph of Fig.~\ref{fig:zq_za_ap2} using the RI/SMOM scheme,
we use the range $a^2p^2\in [2,5]$ for the extrapolation and 
find $Z_q^\msbar/Z_A(2\mbox{ GeV})=1.0842(13)$. The $\chi^2/$dof of this extrapolation is 2.2.
In the RI/SMOM scheme there is no ``democratic" cut on the momenta $p_1$, $p_2$ and $q$. Lattice artefacts proportional
to $a^2p^{[4]}/p^2$ make the data points scatter around the smooth curve in $a^2p^2$ and render the $\chi^2/$dof big.
Thus we enlarge the statistical error from the linear fitting by the factor $\sqrt{\chi^2/{\rm dof}}$ to include this uncertainty. 
In the following analyses for all RCs we similarly inflate the statistical error if the $\chi^2/$dof of the $a^2p^2$ extrapolation
is larger than 1. If using the range $a^2p^2\in[1.5,5]$ or $[2.5,5]$, then we obtain $Z_q^\msbar/Z_A(2\mbox{ GeV})=1.0839(9)$
or 1.0818(20) respectively. The center value changes by 0.0024, which is around $0.2\%$.

Besides the truncation uncertainty in the conversion ratio and the uncertainty from the fitting range of $a^2p^2$
in the linear extrapolation, we also consider the uncertainties from the lattice spacing, 
which is needed to determine the value of $a^2p^2$
corresponding to $p=2$ GeV, from $\Lambda_{\rm QCD}^\msbar$ and from
the perturbative running in the $\msbar$ scheme. For the calculation using RI/MOM as the intermediate scheme,
varying $1/a=1.730(4)$ GeV in one sigma leads to 0.2\% change in $Z_q^\msbar/Z_A(2\mbox{ GeV})$. 
Changing $\Lambda_{\rm QCD}^\msbar=332(17)$ GeV in one sigma leads to 0.1\% change in $Z_q^\msbar/Z_A(2\mbox{ GeV})$.
The perturbative running to 2 GeV of $Z_q^\msbar/Z_A(a^2p^2)$ uses four-loop results of the anomalous dimension. 
The $\mathcal{O}(\alpha_s^4)$ term is found to be around $0.2\%$ of the total size of the running from $a^2p^2>5$ to 2 GeV. 
Similarly we do the analyses for the calculation using RI/SMOM as the intermediate scheme.
The uncertainties of $Z_q^\msbar/Z_A(2\mbox{ GeV})$ are listed in Tab.~\ref{tab:zq_error}.
\begin{table}
\begin{center}
\caption{Uncertainties of $Z_q^\msbar/Z_A(2$ GeV) in the chiral limit. The second and third columns are for using
the RI/MOM and RI/SMOM as the intermediate schemes respectively.}
\begin{tabular}{lcc}
\hline\hline
Source & Error (\%,RI/MOM) & Error (\%,RI/SMOM)  \\
\hline
Statistical & 0.2 & 0.1 \\
\hline
Conversion ratio  & 0.2 & 0.7 \\
$\Lambda_{\rm QCD}^\msbar$  & 0.1 & 0.1 \\
Perturbative running &  0.2 & 0.2 \\
Lattice spacing & 0.2 &  0.1 \\
Fit range of $a^2p^2$ & 0.2 & 0.2 \\
Total sys. uncertainty & 0.4 & 0.8\\
\hline\hline
\end{tabular}
\label{tab:zq_error}
\end{center}
\end{table}

Adding the statistical and systematic uncertainties quadratically, we obtain $Z_q^\msbar/Z_A(2\mbox{ GeV})=1.1027(48)$
and 1.0842(88) respectively for using the RI/MOM and RI/SMOM as the intermediate schemes. 
These two numbers agree with each other within $2\sigma$. Taking 1.1027(48) as our final result and
using the value $Z_A^{\rm WI}=1.1025(9)$ from Sec.~\ref{sec:zawi}, 
we find $Z_q^\msbar(2\mbox{ GeV})=1.2157(54)$ where the error includes
the uncertainty propagated from $Z_A^{\rm WI}$.

\subsection{The tensor operator}
\begin{figure}[htpb]
\centering\includegraphics[width=\textwidth]{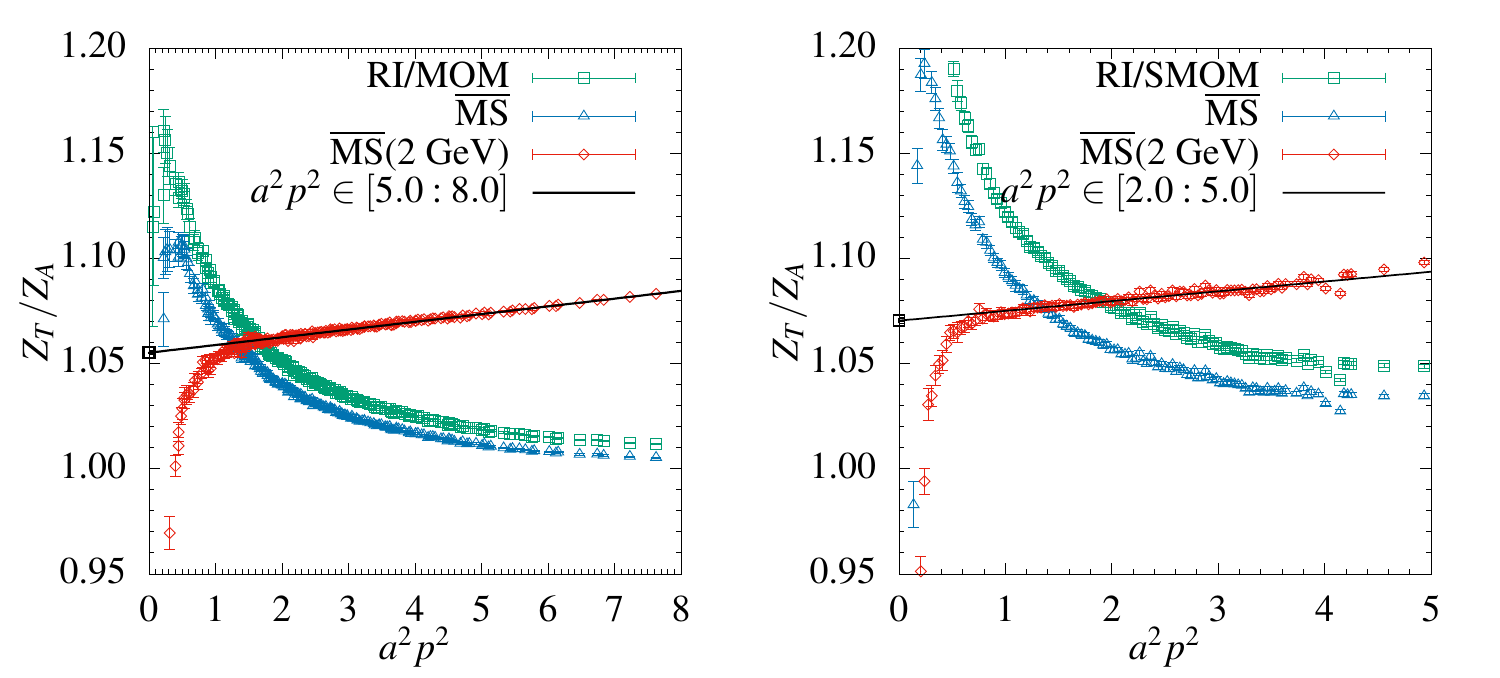}
\caption{Conversion of $Z_T/Z_A$ in the RI/MOM or RI/SMOM scheme to the $\msbar$ scheme. The running to 2 GeV in
the $\msbar$ scheme is shown by the red symbols.}
\label{fig:zt_a2p2}
\end{figure}

From Eq.(\ref{eq:mom}) the ratio $Z_T^{\rm RI/MOM}/Z_A^{\rm WI}=Z_T^{\rm RI/MOM}/Z_A^{\rm RI/MOM}$ 
at a given valence quark mass is computed by
\begin{equation}
\frac{Z_T^{\rm RI/MOM}}{Z_{A}^{\rm RI/MOM}}=\left.\frac{\Gamma_A(p)}{\Gamma_T(p)}\right|_{p^2=\mu^2},
\label{eq:zt_mom}
\end{equation}
where
\begin{equation}
\Gamma_T(p)=\frac{1}{144}\Tr[\Lambda_{T,B}^{\mu\nu}(p)\sigma_{\mu\nu}].
\end{equation}
In the RI/SMOM scheme $Z_T^{\rm RI/SMOM}/Z_A^{\rm WI}(=Z_T^{\rm RI/SMOM}/Z_A^{\rm RI/SMOM})$ at a given valence quark mass is obtained by using
\begin{equation}
\frac{Z_T^{\rm RI/SMOM}}{Z_{A}^{\rm RI/SMOM}}=\left.\frac{\Gamma_A(p_1,p_2)}{\Gamma_T(p_1,p_2)}\right|_{\rm sym},
\label{eq:zt_smom}
\end{equation}
where
\begin{equation}
\Gamma_T(p_1,p_2)=\frac{1}{144}\Tr[\Lambda_{T,B}^{\mu\nu}(p_1,p_2)\sigma_{\mu\nu}].
\end{equation}
The numerical results of both $Z_T^{\rm RI/MOM}/Z_A^{\rm RI/MOM}$ and $Z_T^{\rm RI/SMOM}/Z_A^{\rm RI/SMOM}$ show little
valence quark mass dependence. 
We perform linear extrapolations in $am_q$ to get their chiral limit values, which
are shown by the green squares in Fig.~\ref{fig:zt_a2p2} as functions of the renormalization scale $a^2p^2$.

After the ``democratic" cut in Eq.(\ref{eq:p_cut}) on the momenta, the RI/MOM scheme results show a reasonable smooth behavior in $a^2p^2$.
While some zigzag behavior can be seen in the RI/SMOM scheme results, which is from lattice discretization effects
proportional to $a^2p^{[4]}/p^2$. For the RI/SMOM
scheme, no ``democratic" cut on the momenta $p_1$, $p_2$ and $q$ is applied since 
the symmetric conditions in Eq.(\ref{eq:sym_mom}) are not very easy to satisfy. This leads to the zigzag behavior (especially at
large $a^2p^2$) reflecting
lattice artefacts which are not $O(4)$ invariant.

If we start with the $Z_q^{\rm RI/SMOM}$ defined in Eq.(\ref{eq:zq_smom}), then $Z_T^{\rm RI/SMOM}$ is given by
\begin{equation}
Z_T^{\rm RI/SMOM}=\frac{Z_q^{\rm RI/SMOM}}{\left.\Gamma_T(p_1,p_2)\right|_{\rm sym}}
=\frac{\Tr[S_B^{-1}(p)p\!\!\!/]_{p^2=\mu^2}}{12p^2\left.\Gamma_T(p_1,p_2)\right|_{\rm sym}}.
\label{eq:zt_smom_zq}
\end{equation}
In practice we find that the $Z_T^{\rm RI/SMOM}$ calculated in this way are more scattered around a curve in
$a^2p^2$ indicating larger discretization effects.
In the calculation of the ratio Eq.(\ref{eq:zt_smom}), these $O(4)$ non-invariant effects in the vertex functions in
the denominator and numerator partially cancel. 
The zigzag behavior in the ratio is then less severe.
Thus we always analyze the ratios of other RCs to $Z_A$ and in the end input the value of $Z_A^{\rm WI}$ to obtain their
final results.

The perturbative conversion ratio for $Z_T^{\rm RI/MOM}$ to the $\msbar$ scheme can be obtained from the conversion
ratio for $Z_T^{\rm RI'/MOM}$ to $\msbar$, which was given in Ref.~\cite{Gracey:2003yr} to 3-loops, and
$Z_q^{\rm RI'/MOM}/Z_q^{\rm RI/MOM}$, whose 3-loop result can be computed by using the ratios
$Z_q^{\rm RI'/MOM}/Z_q^\msbar$ and $Z_q^{\rm RI/MOM}/Z_q^\msbar$ given in Ref.~\cite{Chetyrkin:1999pq}. Thus we have
\begin{eqnarray}
&&\frac{Z_T^\msbar}{Z_T^{\rm RI/MOM}}=\frac{Z_T^\msbar}{Z_T^{\rm RI'/MOM}}\frac{Z_T^{\rm RI'/MOM}}{Z_T^{\rm RI/MOM}}=
\frac{Z_T^\msbar}{Z_T^{\rm RI'/MOM}}\frac{Z_q^{\rm RI'/MOM}}{Z_q^{\rm RI/MOM}}\nonumber\\
&=&1-\frac{1}{81}\left(4866-1656\xi_3-259n_f\right)\left(\frac{\alpha_s}{4\pi}\right)^2\nonumber\\
&&+\frac{2}{2187}\left(311424-105984\xi_3-16576n_f\right)\left(\frac{\alpha_s}{4\pi}\right)^3+\mathcal{O}(\alpha_s^4).
\end{eqnarray}
At $p^2=\mu^2=16$ GeV$^2$ and with 3 flavors of dynamical fermions, the numerical value of this ratio is
\begin{eqnarray}
\frac{Z_T^{\msbar}}{Z_T^{\rm RI/MOM}}(\mu=4\mbox{ GeV},n_f=3)&=&1+0.0\alpha_s-0.1641\alpha_s^2+0.0619\alpha_s^3+\cdots\nonumber\\
&=&1-0.0-0.0079+0.0006+\cdots,
\end{eqnarray}
The $\mathcal{O}(\alpha_s^4)$ term is around $0.0004$ if assuming the coefficient of this term is the same as
the one of the $\mathcal{O}(\alpha_s^2)$ term. Thus the truncation error in the conversion ratio is less than $0.1\%$.

The 2-loop matching factor for converting $Z_T^{\rm RI/SMOM}$ to the $\msbar$ scheme is given in Refs.~\cite{Almeida:2010ns,Gracey:2011fb}.
In Landau gauge it reads
\begin{equation}
\frac{Z_T^{\msbar}}{Z_T^{\rm RI/SMOM}}=1-0.21517295\left(\frac{\alpha_s}{4\pi}\right)
-(43.38395007-4.10327859n_f)\left(\frac{\alpha_s}{4\pi}\right)^2+\mathcal{O}(\alpha_s^3).
\end{equation}
For $n_f=3$ and at $\mu=4$ GeV, the above is
\begin{eqnarray}
\frac{Z_T^{\msbar}}{Z_T^{\rm RI/SMOM}}(\mu=4\mbox{ GeV},n_f=3)&=&1-0.017123\alpha_s-0.196779\alpha_s^2+\mathcal{O}(\alpha_s^3)\nonumber\\
&=&1-0.0037-0.0094+\mathcal{O}(\alpha_s^3).
\end{eqnarray}
The 2-loop contribution is larger than the 1-loop contribution and is of size $\sim1\%$.
At $a^2p^2=2$ (or $p=2.447$ GeV) the suppression from $\alpha_s$ is even smaller 
($\alpha_s^\msbar(2.447\mbox{ GeV})=0.2678$).
To be conservative, we assign a $2\%$ truncation error to $Z_T^{\msbar}/Z_T^{\rm RI/SMOM}$ at the scale $p=2.447$ GeV.
Therefore for $Z_T^\msbar$ the conversion uncertainty in using RI/SMOM as the intermediate scheme seems to be much larger than that 
in using the RI/MOM scheme. It will be interesting to really calculate the 3-loop contribution for this conversion ratio.

The blue triangles in both graphs of Fig.~\ref{fig:zt_a2p2} show the ratio $Z_T^\msbar/Z_A$ as a function of the
renormalization scale $a^2p^2$. Their running to 2 GeV is shown by the red diamonds, which are obtained by using the anomalous dimension
$\gamma_T^\msbar$ in Landau gauge calculated up to and including 4-loops. We see a good linear dependence on $a^2p^2$ in
$Z_T^\msbar/Z_A(2\mbox{ GeV};a^2p^2)$ at large $a^2p^2$. Linear extrapolations in $a^2p^2$ in the range $a^2p^2>5$ and $a^2p^2>2$
are done respectively for the results from the two intermediate schemes RI/MOM and RI/SMOM. 
We find $Z_T^\msbar/Z_A(2\mbox{ GeV})=1.0552(5)$ and 1.0704(12) respectively. The ranges are varied to estimate
the associated systematic uncertainties, which are collected in Tab.~\ref{tab:zt_error}.

Similarly to the analyses of the other systematic uncertainties for $Z_q^\msbar/Z_A(2$ GeV) in Sec.~\ref{sec:zq}, we obtain the error budget
for $Z_T^\msbar/Z_A(2$ GeV) in Tab.~\ref{tab:zt_error}.
\begin{table}
\begin{center}
\caption{Uncertainties of $Z_T^\msbar/Z_A(2$ GeV) in the chiral limit}
\begin{tabular}{lcc}
\hline\hline
Source & Error (\%,RI/MOM) & Error (\%,RI/SMOM)  \\
\hline
Statistical & 0.05 & 0.1 \\
\hline
Conversion ratio  & 0.1 & 2 \\
$\Lambda_{\rm QCD}^\msbar$  & 0.1 & 0.1 \\
Perturbative running & $<0.01$ & $<0.01$ \\
Lattice spacing & 0.02 & 0.03 \\
Fit range of $a^2p^2$ & 0.1 & 0.1 \\
Total sys. uncertainty & 0.2 & 2 \\
\hline\hline
\end{tabular}
\label{tab:zt_error}
\end{center}
\end{table}
Adding the statistical and systematic uncertainties quadratically, we get $Z_T^\msbar/Z_A(2\mbox{ GeV})=1.055(2)$
and 1.070(21) respectively for using the RI/MOM and RI/SMOM as the intermediate schemes. 
These two numbers are in agreement within $1\sigma$. The result from the RI/SMOM scheme has a large systematic error from
the conversion ratio.
Taking 1.055(2) as our final result and
using the value $Z_A^{\rm WI}=1.1025(9)$ from Sec.~\ref{sec:zawi}, 
we get $Z_T^\msbar(2\mbox{ GeV})=1.1631(24)$ where the error includes
the uncertainty propagated from $Z_A^{\rm WI}$.

\subsection{The scalar density}\label{sec:zs}
The calculation of $Z_S^{\rm RI/MOM}$ and its conversion to the $\msbar$ scheme closely follow our previous work~\cite{Liu:2013yxz}
on the $24^3\times64$ RBC/UKQCD lattices with similar lattice spacings. 
One difference is that now we can impose a stronger ``democratic" cut (Eq.~\ref{eq:p_cut}) 
on the momentum modes since
the lattice size is now bigger. This leads to a smoother dependence on the renormalization scale $a^2p^2$ in $Z_S^{\rm RI/MOM}$ since the
Lorentz non-invariant lattice artefacts are further reduced. Another difference is that here we analyze the ratio 
$Z_S^{\rm RI/MOM}/Z_A^{\rm RI/MOM}=\Gamma_A(p)/\Gamma_S(p)|_{p^2=\mu^2}$ instead of the absolute $Z_S^{\rm RI/MOM}$.

\begin{figure}[htpb]
\centering\includegraphics[width=\textwidth]{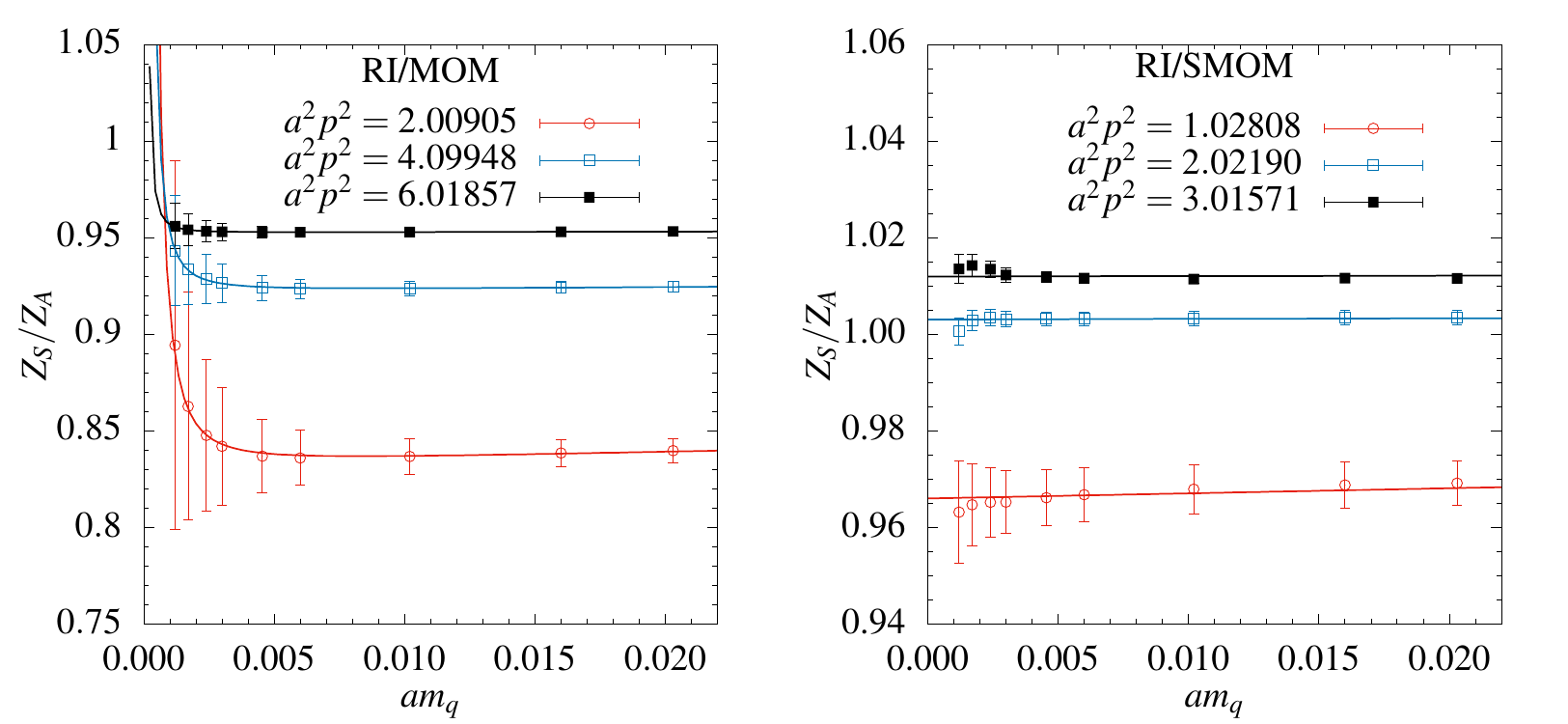}
\caption{Chiral extrapolations of $Z_S/Z_A$ in the RI/MOM scheme using Eq.(\ref{eq:zs_fit}) (left panel) and in the RI/SMOM scheme using a linear function (right panel).}
\label{fig:zs_chiral}
\end{figure}

The chiral extrapolation of $Z_S^{\rm RI/MOM}/Z_A^{\rm RI/MOM}$ is done with an ansatz 
with three parameters $A_s$, $B_s$ and $C_s$~\cite{Aoki:2007xm,Blum:2001sr,Liu:2013yxz}
\begin{equation}
Z_S/Z_A=\frac{A_s}{(am_q)^2}+B_s+C_s\cdot am_q,
\label{eq:zs_fit}
\end{equation}
where the double pole term comes from the topological zero modes of the overlap fermions. $B_s$ is taken as the chiral limit value
of $Z_S^{\rm RI/MOM}/Z_A^{\rm RI/MOM}$.
Examples of this extrapolation are shown in the left panel of Fig.~\ref{fig:zs_chiral}.

The valence quark mass dependence of $Z_S^{\rm RI/SMOM}/Z_A^{\rm RI/SMOM}$ seems to be milder,
which is computed from the ratio $\Gamma_A(p_1,p_2)/\Gamma_S(p_1,p_2)|_{\rm sym}$.
We tried both Eq.(\ref{eq:zs_fit}) and a linear function in quark mass
for going to the chiral limit (only the linear extrapolation is shown in the right panel of Fig.~\ref{fig:zs_chiral}). 
We find consistent chiral limit values
from the two extrapolations for $Z_S^{\rm RI/SMOM}/Z_A^{\rm RI/SMOM}$.

\begin{figure}[htpb]
\centering\includegraphics[width=\textwidth]{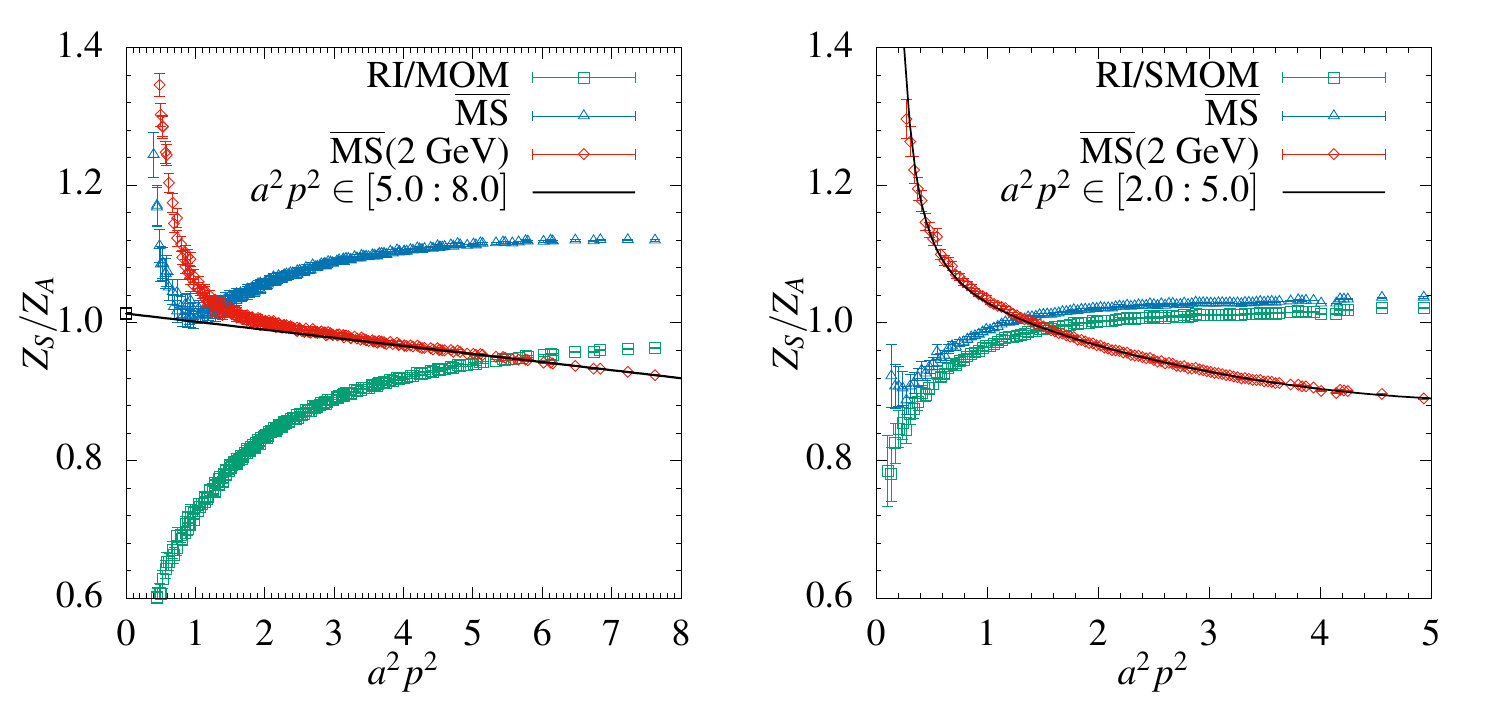}
\caption{Conversion and running of $Z_S/Z_A$ for the intermediate schemes RI/MOM (left) and RI/SMOM (right).
The black curve in the right panel is a fitting using ansatz (e) in Eq.(\ref{eq:fx5}).}
\label{fig:zs_running}
\end{figure}

The scale dependence of $Z_S/Z_A$ is shown in Fig.~\ref{fig:zs_running} by the green squares.
The conversion to $\msbar$ and the running to 2 GeV in $\msbar$ are shown in the same figure by the blue triangles and the red diamonds respectively.
In the right panel the green squares and the blue triangles are much closer together than in the left panel. This is because
the conversion ratio from the RI/SMOM scheme to the $\msbar$ scheme is much closer to $1$ than the ratio from the RI/MOM
scheme to $\msbar$. This leads to a much smaller systematic uncertainty from the conversion
in using RI/SMOM as the intermediate scheme. As we have estimated in our previous work~\cite{Liu:2013yxz},
the truncation error in the conversion ratio $Z_S^{\msbar}/Z_S^{\rm RI/MOM}$ (3-loop result) above 4 GeV is around $1.5\%$.
The inverse of the conversion ratio $Z_S^{\msbar}/Z_S^{\rm RI/SMOM}$ have been calculated up to two loops~\cite{Gorbahn:2010bf,Almeida:2010ns}
\begin{equation}
\frac{Z_S^{\rm RI/SMOM}}{Z_S^{\msbar}}=\frac{Z_m^{\msbar}}{Z_m^{\rm RI/SMOM}}=1-0.6455188560\left(\frac{\alpha_s}{4\pi}\right)
-(22.60768757-4.013539470n_f)\left(\frac{\alpha_s}{4\pi}\right)^2.
\end{equation}
From the above we find for $n_f=3$ and at $a^2p^2=2$ ($p=2.447$ GeV)
\begin{eqnarray}
\frac{Z_S^{\msbar}}{Z_S^{\rm RI/SMOM}}&=&1.0+0.051369\alpha_s+0.069556\alpha_s^2+\mathcal{O}(\alpha_s^3)\nonumber\\
&\stackrel{p=2.447\mbox{ \tiny{GeV}}}{=}&1.0+0.0138+0.0050+\mathcal{O}(\alpha_s^3).
\end{eqnarray}
Assuming the coefficient for the $\mathcal{O}(\alpha_s^3)$ term is $0.069556\times(0.069556/0.051369)=0.094182$, we
expect the $\mathcal{O}(\alpha_s^3)$ term to be of size $\sim0.0018$. This means the truncation error can be estimated
to be $0.2\%$, which is much smaller than the $1.5\%$ for the conversion of the RI/MOM scheme result.

After the running to 2 GeV in the $\msbar$ scheme,  $Z_S^{\msbar}(2\mbox{ GeV}; a^2p^2)/Z_A$ shows a good linear behavior in 
the initial scale $a^2p^2$
at large scales in using RI/MOM as the intermediate scheme. 
This is shown by the red diamonds in the left panel of Fig.~\ref{fig:zs_running}.
The nonzero slope of the red diamonds is attributed to lattice discretization effects proportional to $a^2p^2$.
Above $a^2p^2=5$ we can do a linear extrapolation to $a^2p^2=0$ with good $\chi^2/$dof and obtain $Z_S^{\msbar}(2\mbox{ GeV})/Z_A=1.0137(13)$.
Varying the extrapolation range to $a^2p^2>4$, we get $Z_S^{\msbar}(2\mbox{ GeV})/Z_A=1.0152(8)$. The variation
in the center value (0.1\%) is taken as one of the systematic uncertainties.
\begin{table}
\begin{center}
\caption{Uncertainties of $Z_S^\msbar/Z_A(2$ GeV) and $Z_P^\msbar/Z_A(2$ GeV) in the chiral limit through the RI/MOM scheme.}
\begin{tabular}{lcc}
\hline\hline
Source & $Z_S^\msbar/Z_A(2$ GeV) (\%) & $Z_P^\msbar/Z_A(2$ GeV) (\%)  \\
\hline
Statistical & 0.1 & 0.6 \\
\hline
Conversion ratio  & 1.5 & 1.5 \\
$\Lambda_{\rm QCD}^\msbar$  & 0.3 & 0.4 \\
Perturbative running &  0.1 & 0.1 \\
Lattice spacing & 0.1 &  $<0.1$ \\
Fit range of $a^2p^2$ & 0.1 & 0.3 \\
Total sys. uncertainty & 1.6 & 1.6 \\
\hline\hline
\end{tabular}
\label{tab:zs_error}
\end{center}
\end{table}

The statistical and systematic uncertainties of $Z_S^\msbar/Z_A(2$ GeV) obtained by using RI/MOM as the intermediate scheme
are listed in Tab.~\ref{tab:zs_error}. The uncertainty from the conversion ratio dominates.
Adding all the uncertainties quadratically, we get $Z_S^\msbar/Z_A(2\mbox{ GeV})=1.014(16)$. 
Using the value $Z_A^{\rm WI}=1.1025(9)$ from Sec.~\ref{sec:zawi}, we find $Z_S^\msbar(2\mbox{ GeV})=1.118(18)$ where the error includes
the uncertainty propagated from $Z_A^{\rm WI}$. 
This number agrees with our previous result $Z_S^\msbar(2\mbox{ GeV})=1.127(21)$ on the $24^3\times64$ lattice~\cite{Liu:2013yxz}.

In the right graph of Fig.~\ref{fig:zs_running}, $Z_S^{\msbar}(2\mbox{ GeV}; a^2p^2)/Z_A$ obtained by using RI/SMOM
as the intermediate scheme is shown as a function of the initial scale $a^2p^2$. 
We do not see a clear window, in which $Z_S^{\msbar}(2\mbox{ GeV}; a^2p^2)/Z_A$ linearly depends on $a^2p^2$.
This is quite different from what we saw in $Z_q^{\msbar}(2\mbox{ GeV}; a^2p^2)/Z_A$ and $Z_T^{\msbar}(2\mbox{ GeV}; a^2p^2)/Z_A$
(the right panels in Figs.~\ref{fig:zq_za_ap2} and~\ref{fig:zt_a2p2} respectively), which are also obtained through the RI/SMOM scheme. 
To model the behavior of $Z_S^{\msbar}(2\mbox{ GeV}; a^2p^2)/Z_A$, we tried several ansatzes 
(with $x\equiv a^2p^2$ and fitting parameters $A$, $B$, $C$, $D$ and $E$):
\begin{eqnarray}
(a):\quad f(x)&=&A+Bx+Cx^2,\label{eq:fx1}\\
(b):\quad f(x)&=&A+Bx+D/x,\label{eq:fx2}\\
(c):\quad f(x)&=&A+Bx+E/x^2,\label{eq:fx3}\\
(d):\quad f(x)&=&A+Bx+Cx^2+D/x,\label{eq:fx4}\\
(e):\quad f(x)&=&A+Bx+Cx^2+E/x^2,\label{eq:fx5}
\end{eqnarray}
where the $C (a^2p^2)^2$ term is for higher order discretization effects and the $1/x^n$ ($n=1,2$) terms are for
possible nonperturbative effects.
In using the above ansatzes to fit our data, we fix the upper limit of $a^2p^2$ to $5$ and vary the lower limit.
We collect the $\chi^2/$dof and the results of $A$ for various fitting ranges in Tab.~\ref{tab:zs_fits}.
\begin{table}
\begin{center}
\caption{Fittings of $Z_S^\msbar/Z_A(2$ GeV;$a^2p^2$) obtained from the RI/SMOM scheme to the models in Eqs.(\ref{eq:fx1}-\ref{eq:fx5}).
Here we vary the lower limit of $x(\equiv a^2p^2)$. The uncertainties have been inflated by the factor $\sqrt{\chi^2/{\rm dof}}$.}
{\footnotesize
\begin{tabular}{c|cc|cc|cc|cc|cc}
\hline\hline
 & \multicolumn{2}{c|}{$(a)$} & \multicolumn{2}{c|}{$(b)$} & \multicolumn{2}{c|}{$(c)$} & \multicolumn{2}{c|}{$(d)$} & \multicolumn{2}{c}{$(e)$} \\
 \cline{2-11}
$x_{\rm min}$ & $A$  & $\chi^2/\mbox{dof}$  & $A$  &  $\chi^2/$dof & $A$ & $\chi^2/\mbox{dof}$ & $A$ & $\chi^2/\mbox{dof}$   & $A$   & $\chi^2/\mbox{dof}$ \\ 
\hline
	0.5  & 1.096(2) & 8.56 & 0.941(3)   & 7.83      & 1.000(2) & 20.2 & 1.015(6)   & 2.95       & 1.064(3)    & 3.01 \\
	1.0  & 1.091(2) & 5.37 & 0.930(4)   & 6.79      & 0.986(2) & 12.2 & 1.020(10)  & 3.41       & 1.057(5)    & 3.41 \\
	1.5  & 1.084(2) & 4.52 & 0.915(5)   & 6.10      & 0.973(3) & 8.57 & 1.027(19)  & 4.03       & 1.056(9)    & 4.02             \\
	1.8  & 1.080(3) & 4.16 & 0.904(6)   & 6.07      & 0.964(3) & 7.58 & 1.040(30)  & 4.54       & 1.060(15)   & 4.54             \\
	2.0  & 1.078(3) & 4.92 & 0.895(7)   & 6.13      & 0.957(4) & 7.19 & 1.050(42)  & 4.97       & 1.064(21)   & 4.97             \\
	2.2  & 1.077(4) & 5.39 & 0.883(9)   & 6.18      & 0.948(4) & 5.47 & 1.046(61)  & 5.47       & 1.061(30)   & 5.47             \\
	2.5  & 1.074(6) & 5.99 & 0.869(14)  & 6.63      & 0.938(7) & 7.04 & 1.08(10)   & 6.14       & 1.078(53)   & 6.14             \\
	2.8  & 1.074(10)& 6.69 & 0.843(22)  & 6.98      & 0.921(12)& 7.18 & 1.08(20)   & 6.89       & 1.08(10)    & 6.89            \\
	3.0  & 1.067(15)& 7.41 & 0.834(33)  & 7.68      & 0.913(17)& 7.81 & 1.17(33)   & 7.66       & 1.12(17)    & 7.66            \\
\hline\hline
\end{tabular}
}
\label{tab:zs_fits}
\end{center}
\end{table}
We find the $C (a^2p^2)^2$ term is necessary to decrease the $\chi^2/$dof. Models $(a)$, $(d)$ and $(e)$ give smaller $\chi^2/$dof
than models $(b)$ and $(c)$ in almost all fitting ranges (the only exception case is with $(a^2p^2)_{\rm min}=0.5$, in which model $(a)$ gives
a larger $\chi^2/$dof than model $(b)$).
For the behavior of the possible nonperturbative effects, the data can hardly distinguish between $1/x$ and $1/x^2$ 
(the $\chi^2/$dof's of models $(d)$ and $(e)$ are almost all the same for all fitting ranges).
Above $(a^2p^2)_{\rm min}=\sim1.5$, the possible nonperturbative effects ($1/x^n$ terms) can be ignored: Model $(a)$ gives
comparable $\chi^2/$dof as or smaller $\chi^2/$dof than models $(d)$ and $(e)$; 
Also the fitted parameter $D$($E$) in model $(d)$($(e)$) is consistent with zero within its uncertainty. 
This is of course expected since the nonperturbative
effects are suppressed at large momentum scale.

\begin{figure}[htpb]
\centering\includegraphics[width=0.5\textwidth]{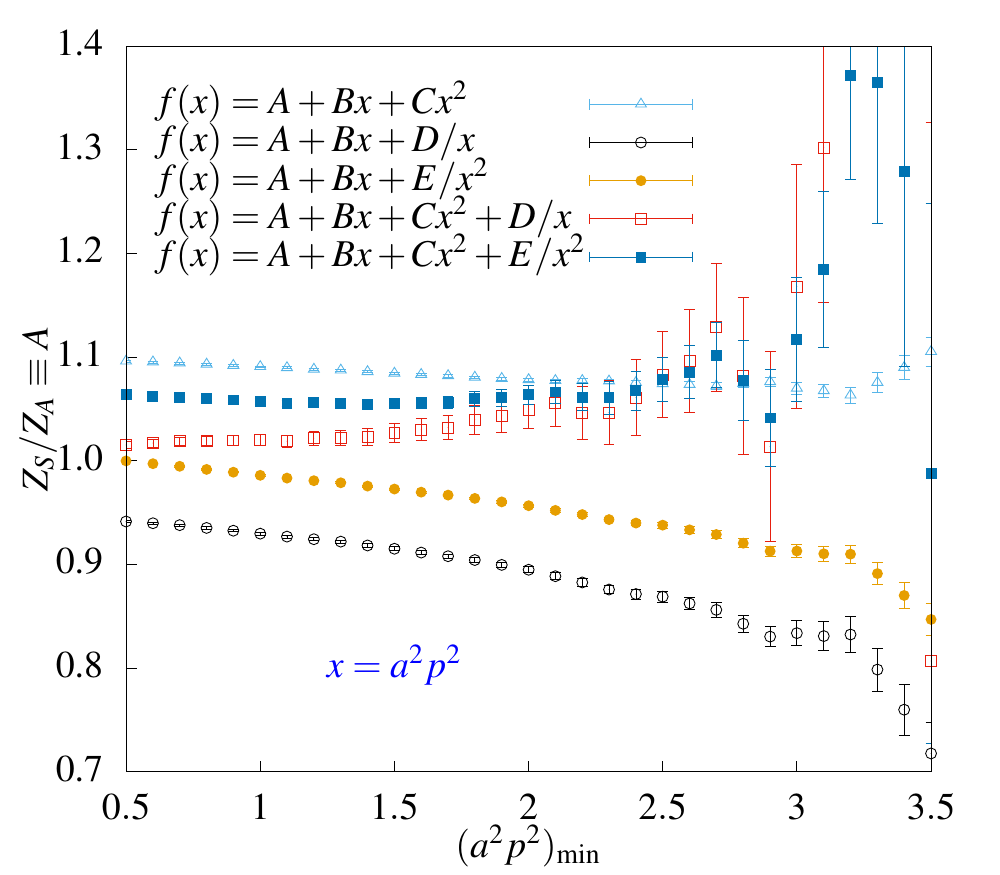}
\caption{$Z_S^\msbar/Z_A(2\mbox{ GeV})$ from the RI/SMOM scheme by fitting $Z_S^\msbar/Z_A(2$ GeV;$a^2p^2$) to the five ansatzes Eqs.(\ref{eq:fx1}-\ref{eq:fx5}) as we vary
the lower limit $(a^2p^2)_{\rm min}$ of the fitting range $[(a^2p^2)_{\rm min}, 5]$.}
\label{fig:zs_models}
\end{figure}

The resulted $A$'s from all fittings are plotted in Fig.~\ref{fig:zs_models} as functions of the lower limit $(a^2p^2)_{\rm min}$.
Above $(a^2p^2)_{\rm min}=\sim 2$, the results from models $(a)$, $(d)$ and $(e)$ converge to a more or less stable value.
In the range $(a^2p^2)_{\rm min}\in [2, 3]$ the three models give consistent results. 
Taking $(a^2p^2)_{\rm min}=2$ and averaging the $A$'s from the three models, we find $Z_S^\msbar/Z_A(2\mbox{ GeV})=1.064(42)$.
Here the statistical error is the biggest of the three ones from the three fittings. 
The span (0.028 or 2.6\%) in the three center values will be
taken as one source of the systematic uncertainties. 

In Tab.~\ref{tab:zsp_smom_error} we collect our error analyses for
$Z_S^\msbar/Z_A(2\mbox{ GeV})$ obtained through the RI/SMOM scheme.
\begin{table}
\begin{center}
\caption{Uncertainties of $Z_S^\msbar/Z_A(2$ GeV) and $Z_P^\msbar/Z_A(2$ GeV) in the chiral limit through the RI/SMOM scheme.}
\begin{tabular}{lcc}
\hline\hline
Source & $Z_S^\msbar/Z_A(2$ GeV) (\%) & $Z_P^\msbar/Z_A(2$ GeV) (\%)  \\
\hline
Statistical & 3.9 &  4.4 \\
\hline
Conversion ratio  & 0.2 & 0.2 \\
$\Lambda_{\rm QCD}^\msbar$  & 0.2 & 0.1 \\
Perturbative running & 0.1 & 0.1 \\
Lattice spacing & 0.1 &  $<0.1$ \\
Fit range of $a^2p^2$ & 2.1 & 1.2 \\
Span in the results from models $(a)$, $(d)$ \& $(e)$ & 2.6 & 3.0 \\
Total sys. uncertainty & 3.4 & 3.2 \\
\hline\hline
\end{tabular}
\label{tab:zsp_smom_error}
\end{center}
\end{table}
In total, we find a $5.1\%$ error in $Z_S^\msbar/Z_A(2$ GeV). That is to say, $Z_S^\msbar/Z_A(2\mbox{ GeV})=1.064(55)$.
This result, with a relatively large error, agrees with $Z_S^\msbar/Z_A(2\mbox{ GeV})=1.014(16)$ obtained by using the RI/MOM scheme.
The fact that we cannot get a broad window of $a^2p^2$ in $Z_S^\msbar/Z_A(2\mbox{ GeV}; a^2p^2)$, 
in which both the nonperturbative effects and the lattice discretization effects are small, 
leads to the large uncertainty in $Z_S^\msbar/Z_A(2$ GeV). This may come from the HYP smearing that we do on the gauge fields.
In Ref.~\cite{Arthur:2013bqa} the upper edge of the renormalization window was found to be reduced by link smearing.
We may do a calculation on thin link configurations to check this in the future.

\subsection{The pseudoscalar density}
\begin{figure}[htpb]
\centering\includegraphics[width=\textwidth]{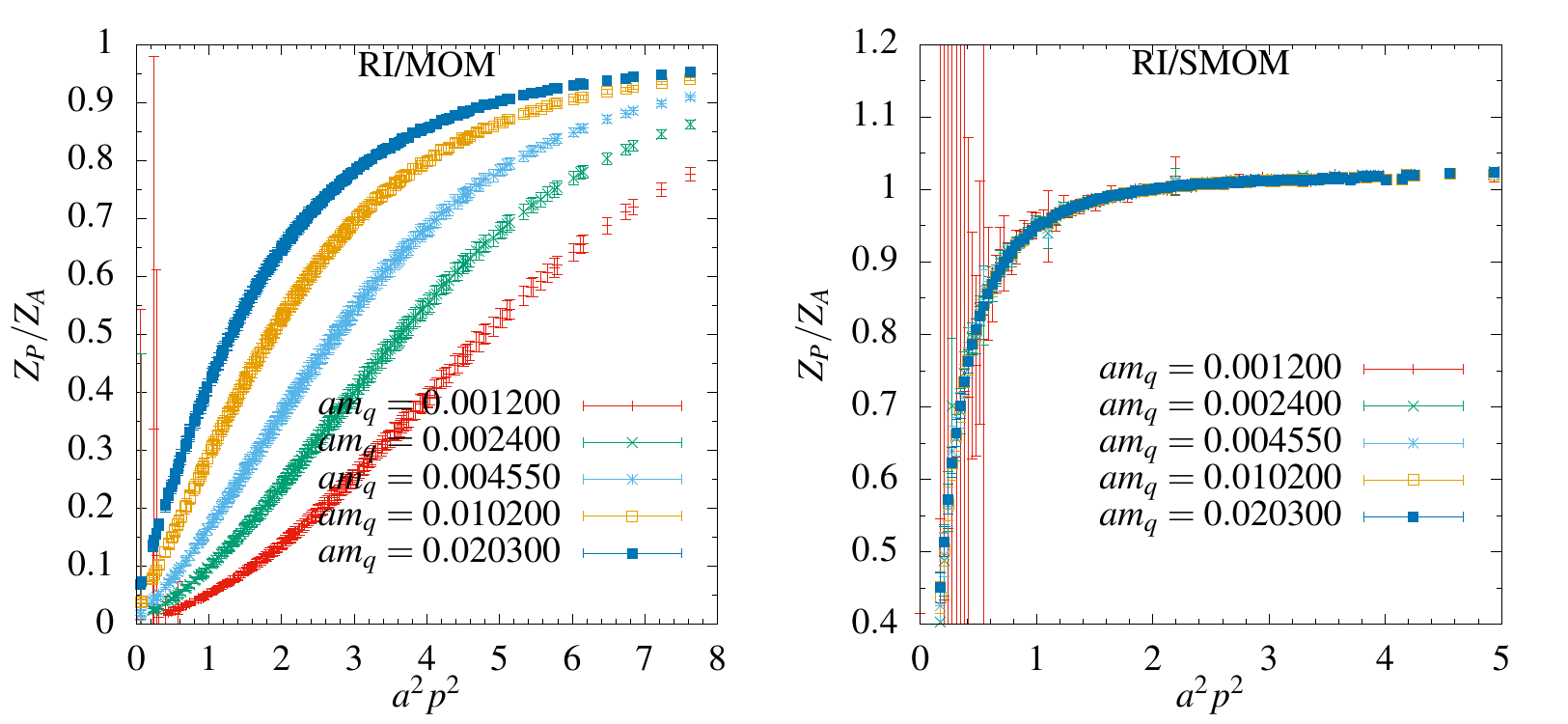}
\caption{$Z_P/Z_A$ in the schemes RI/MOM (left graph) and RI/SMOM (right graph).}
\label{fig:zp_raw}
\end{figure}

\begin{figure}[htpb]
\centering\includegraphics[width=\textwidth]{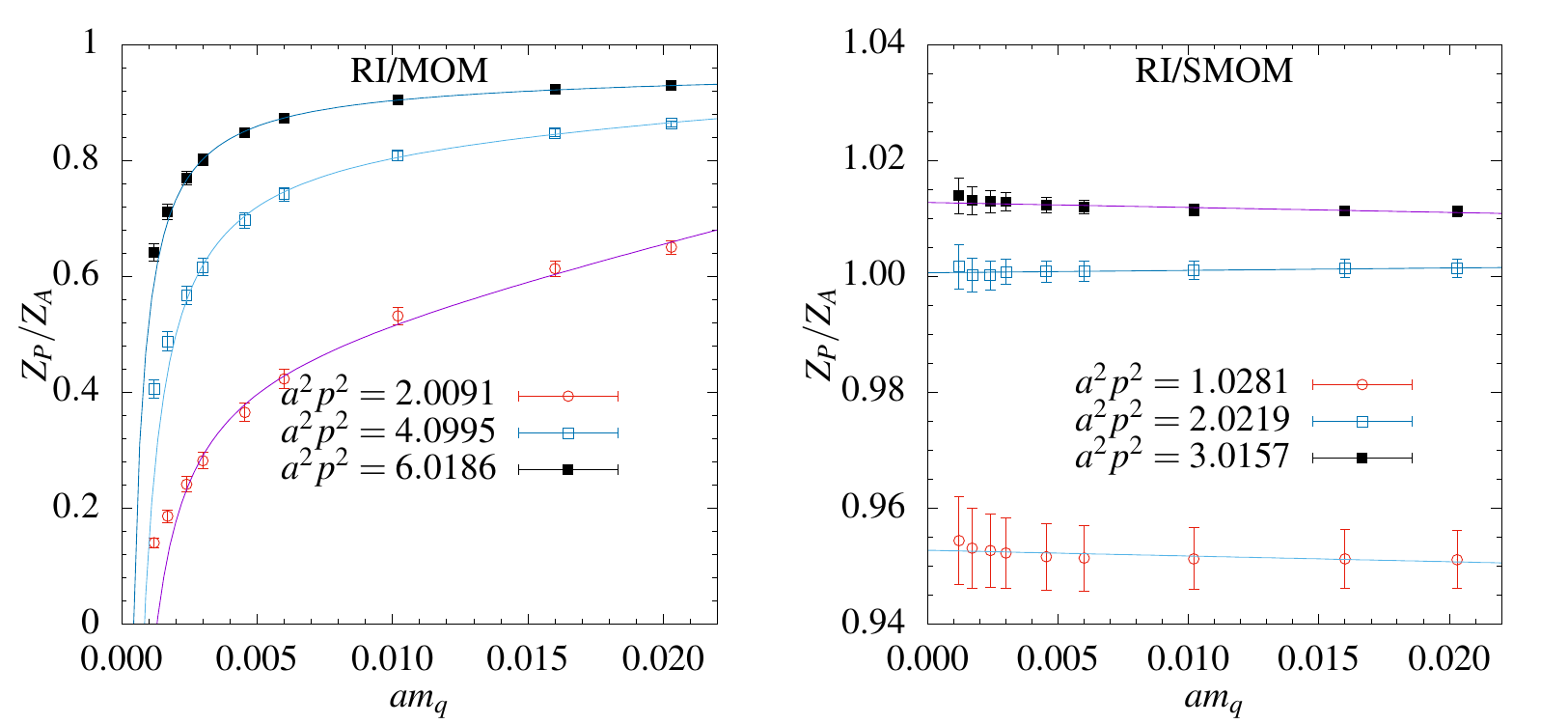}
\caption{Examples of chiral extrapolations of $Z_P/Z_A$ in the RI/MOM (left graph) and RI/SMOM (right graph) schemes.}
\label{fig:zp_chiral}
\end{figure}

For the calculation of $Z_P^{\rm RI/MOM}/Z_A$ and its conversion to $\msbar$, we again closely follow our previous work~\cite{Liu:2013yxz}.
The Goldstone boson contamination in $Z_P^{\rm RI/MOM}/Z_A$ is apparent as we can see from the left panel in Fig.~\ref{fig:zp_raw}.
This contamination in the forward vertex function $\Gamma_P(p)$ is proportional to $1/(m_\pi^2 p^2)$ or $1/(m_q p^2)$ due to
the pion propagator. 
Thus the chiral extrapolation of $Z_P^{\rm RI/MOM}/Z_A$ at a fixed $a^2p^2$ is done by using the ansatz
\begin{equation}
(Z_P^{\rm RI/MOM}/Z_A)^{-1}=\frac{A_p}{am_q}+B_p+C_p\cdot am_q,
\label{eq:zp_chiral}
\end{equation}
where $A_p, B_p$ and $C_p$ are three fitting parameters. $B_p^{-1}$ is taken as the chiral limit value for $Z_P^{\rm RI/MOM}/Z_A$.
Examples of this chiral extrapolation are shown in the left panel of Fig.~\ref{fig:zp_chiral}.

In the SMOM scheme, the Goldstone boson contamination in $\Gamma_P(p_1,p_2)|_{\rm sym}$ is proportional to $1/q^2(=1/p^2)$ 
since $m_\pi^2\ll q^2$.
Thus the quark mass dependence of $Z_P^{\rm RI/SMOM}/Z_A$ should be small providing $m_q^2\ll q^2$.
This can be clearly seen in the right graph of Fig.~\ref{fig:zp_raw} and also in the right panel of Fig.~\ref{fig:zp_chiral},
in which we do linear chiral extrapolations for $Z_P^{\rm RI/SMOM}/Z_A$ at three typical momentum values.

\begin{figure}[htpb]
\centering\includegraphics[width=0.5\textwidth]{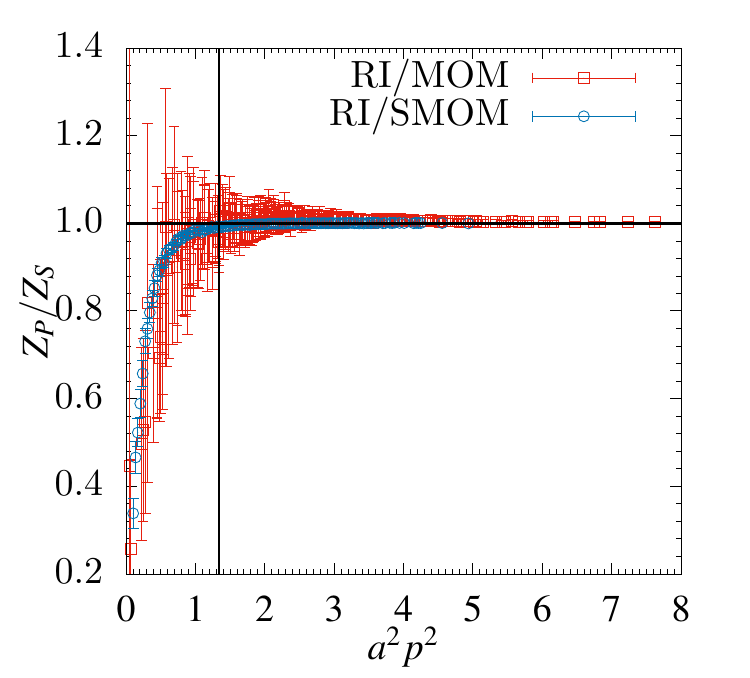}
\caption{$Z_P/Z_S$ in the extrapolated chiral limit in the RI/MOM and RI/SMOM schemes. $Z_P/Z_S=1$ is well satisfied in both schemes at large momentum scale.
In the RI/MOM result the Goldstone boson contamination has been removed by
using Eq.(\ref{eq:zp_chiral}). The horizontal line is $Z_P/Z_S=1$ for guiding the eyes. The vertical line indicates $p=2$ GeV.}
\label{fig:zsp}
\end{figure}

Fig.~\ref{fig:zsp} shows
the ratio $Z_P/Z_S$ in the extrapolated chiral limit in both the RI/MOM and RI/SMOM schemes.
These results verify the relation $Z_S=Z_P$ for overlap fermions. The RI/SMOM scheme suppresses the Goldstone boson contamination
in the pseudoscalar vertex function since $q^2\neq0$. Thus $Z_P/Z_S=1$ is more precisely satisfied in this scheme than in the
RI/MOM scheme.

\begin{figure}[htpb]
\centering\includegraphics[width=\textwidth]{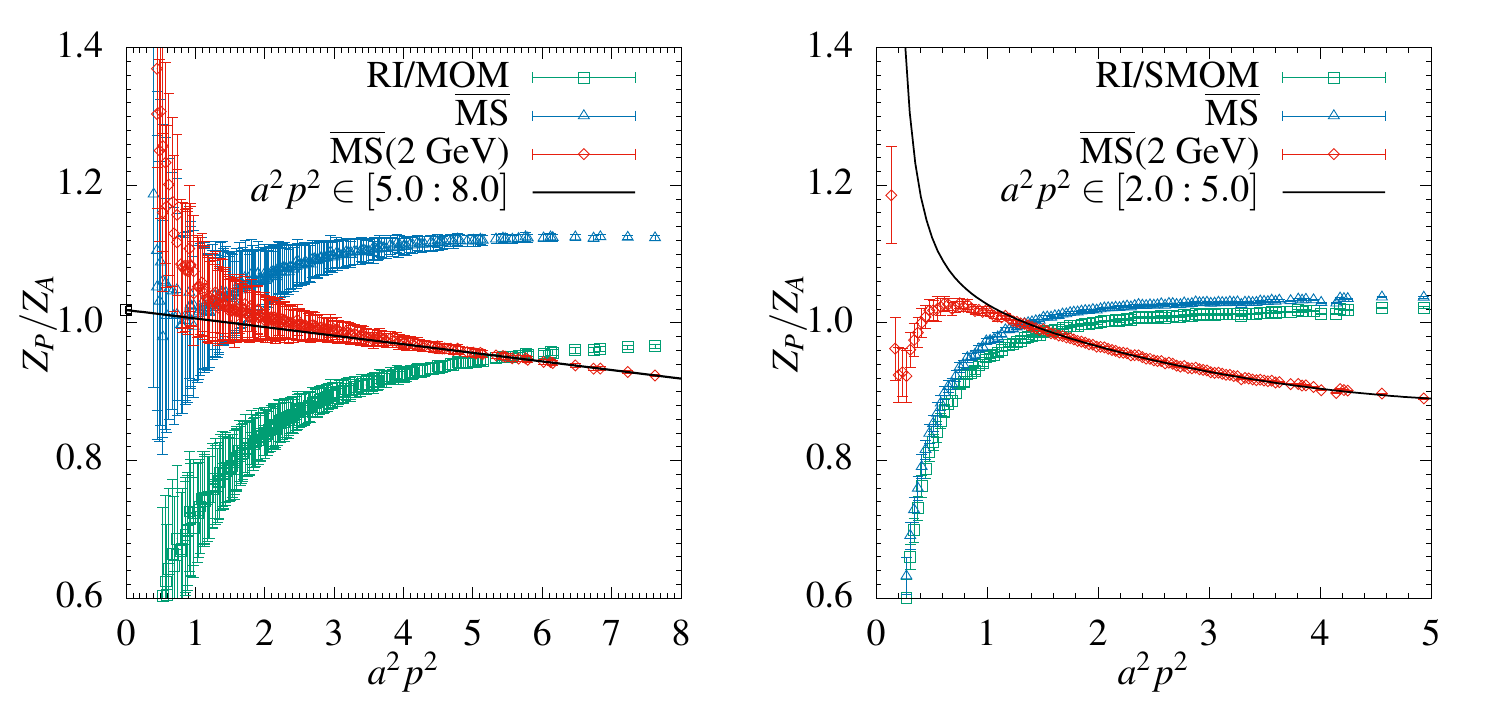}
\caption{The conversion and running of $Z_P/Z_A$ using RI/MOM (left) or RI/SMOM (right) as
the intermediate scheme. The black curve in the right panel is a fitting using ansatz (e) in Eq.(\ref{eq:fx5}).}
\label{fig:zp_running}
\end{figure}

The conversion to and running in the $\msbar$ scheme for both $Z_P^{\rm RI/MOM}/Z_A$ and $Z_P^{\rm RI/SMOM}/Z_A$
are shown in Fig.~\ref{fig:zp_running}. The two conversion ratios used here are the same as those for
$Z_S^{\rm RI/MOM}/Z_A$ and $Z_S^{\rm RI/SMOM}/Z_A$ respectively.
From the RI/MOM result, a linear extrapolation in $a^2p^2$ for $Z_P^{\msbar}(2\mbox{ GeV}; a^2p^2)/Z_A$ gives
$Z_P^{\msbar}(2\mbox{ GeV})/Z_A=1.019(6)$. It agrees with $Z_S^{\msbar}(2\mbox{ GeV})/Z_A=1.0137(13)$ obtained 
in Sec.~\ref{sec:zs} also by using
RI/MOM as the intermediate scheme. 
In the last column of Table~\ref{tab:zs_error} we give our error analyses for $Z_P^{\msbar}(2\mbox{ GeV})/Z_A$.
Our final result is $Z_P^{\msbar}(2\mbox{ GeV})/Z_A=1.019(6)(16)$, where the second error is the total systematic uncertainty.
Using the value $Z_A^{\rm WI}=1.1025(9)$ from Sec.~\ref{sec:zawi}, 
we get $Z_P^\msbar(2\mbox{ GeV})=1.123(19)$ where the error includes
the uncertainty propagated from $Z_A^{\rm WI}$.

When we convert the RI/SMOM result $Z_P^{\rm RI/SMOM}(a^2p^2)/Z_A$ into the $\msbar$ scheme and run it to 2 GeV, 
we find that the behavior of $Z_P^{\msbar}(2\mbox{ GeV}; a^2p^2)/Z_A$ 
(see the right panel in Fig.~\ref{fig:zp_running})
looks similar to that of $Z_S^{\msbar}(2\mbox{ GeV}; a^2p^2)/Z_A$ in the right panel of Fig.~\ref{fig:zs_running}
except at the very small $a^2p^2$ region.
We therefore also tried the ansatzes in Eqs.(\ref{eq:fx1})-(\ref{eq:fx5}) to remove the lattice discretization artefacts and
possible nonperturbative effects in $Z_P^{\msbar}(2\mbox{ GeV}; a^2p^2)/Z_A$ to get $Z_P^{\msbar}(2\mbox{ GeV})/Z_A$.
The fitting results are similar to those for $Z_S^{\msbar}(2\mbox{ GeV}; a^2p^2)/Z_A$ from the RI/SMOM scheme:
The $C(a^2p^2)^2$ term is necessary to reduce the $\chi^2/$dof; 
Above $(a^2p^2)_{\rm min}=\sim1.5$, the nonperturbative effects ($1/x^n$ terms) can be ignored.

\begin{figure}[htpb]
\centering\includegraphics[width=0.5\textwidth]{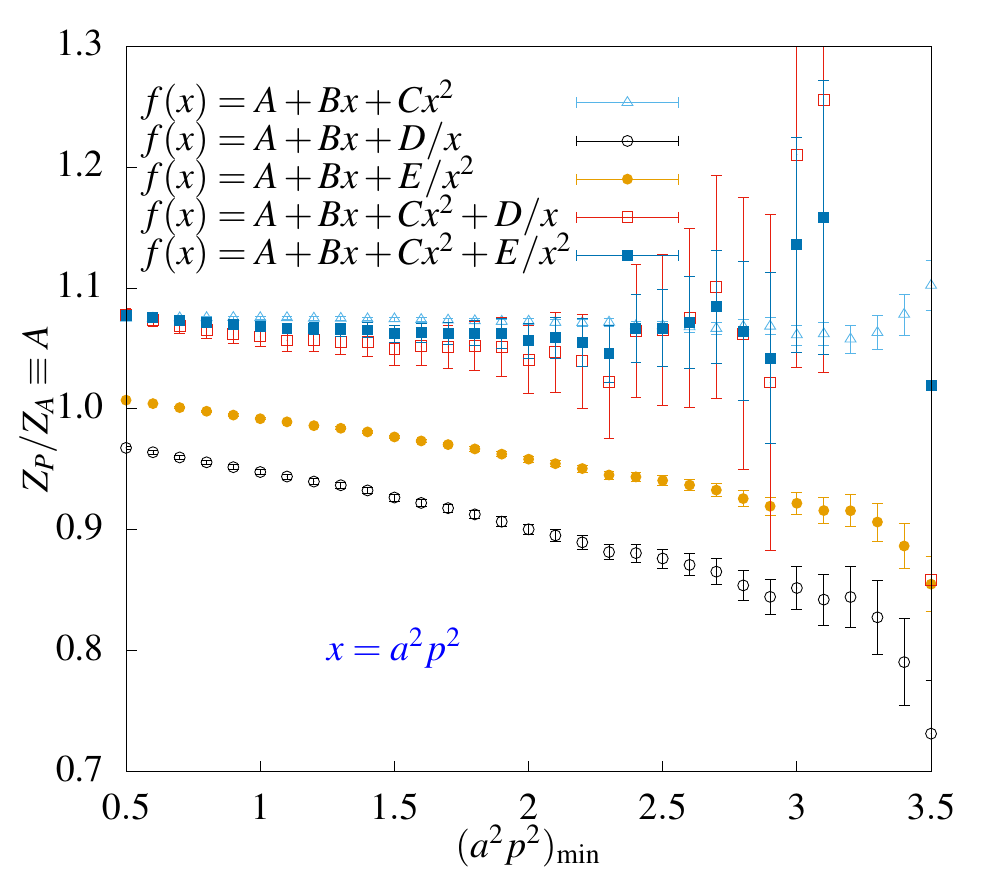}
\caption{Results of $A$ from fitting $Z_P^\msbar/Z_A(2$ GeV;$a^2p^2$) to the ansatzes in Eqs.(\ref{eq:fx1})-(\ref{eq:fx5})
when using RI/SMOM as the intermediate scheme. The lower limit $(a^2p^2)_{\rm min}$ of the fitting range $[(a^2p^2)_{\rm min}, 5]$
is varied.}
\label{fig:zp_models}
\end{figure}

A graph similar to Fig.~\ref{fig:zs_models} is obtained as shown in Fig.~\ref{fig:zp_models} for the fitted parameter 
$A=Z_P^{\msbar}(2\mbox{ GeV})/Z_A$.
Averaging the results from models $(a)$, $(d)$ and $(e)$ with $(a^2p^2)_{\rm min}=2$, we get $Z_P^\msbar(2\mbox{ GeV})/Z_A=1.056(47)$.
Here again we take the biggest uncertainty from the three fits as our statistical error.
The analyses of uncertainties for $Z_P^\msbar(2\mbox{ GeV})/Z_A$ are given in Table~\ref{tab:zsp_smom_error}.
Adding all the uncertainties quadratically, we finally obtain $Z_P^\msbar(2\mbox{ GeV})/Z_A=1.056(58)$.
This number agrees with $Z_S^\msbar(2\mbox{ GeV})/Z_A=1.064(55)$ from using also the RI/SMOM scheme.

\section{Summary}
\label{sec:summary}
In this paper we present our calculation of the RCs for the quark field ($Z_q$) and 
bilinear quark operators ($Z_S$, $Z_P$, $Z_V$, $Z_A$ and $Z_T$) by using the
RI/MOM and RI/SMOM schemes. Our lattice setup is overlap valence quark on 2+1-flavor domain wall fermion configurations
generated by the RBC-UKQCD Collaborations at the physical pion mass ($m_\pi^{\rm sea}=139.2(4)$ MeV)~\cite{Blum:2014tka}. 
The lattices are of size $48^3\times96$, on which
the $\chi$QCD Collaboration is studying many physical quantities such as vector meson decay constants.
We compute $Z_A$ from the PCAC relation and obtain the ratios of other RCs to $Z_A$
from the appropriate vertex functions. The results are converted to the $\msbar$ scheme and the scale is
set to 2 GeV when there is a scale dependence. These matching factors are necessary to connect the lattice results to
the continuum world. For the convenience of later usage we collect our final results in Table~\ref{tab:z_final},
where the uncertainties include both the statistical and systematic errors.
\begin{table}
\begin{center}
\caption{Matching factors to the $\msbar$ scheme for the quark field and bilinear quark operators.}
\begin{tabular}{ccccc}
\hline\hline
 $Z_A$     & $Z_q(2\mbox{ GeV})$ & $Z_T(2\mbox{ GeV})$ & $Z_S(2\mbox{ GeV})$ & $Z_P(2\mbox{ GeV})$ \\
 1.1025(9) &   1.2157(54)        &   1.1631(24)        & 1.118(18)           & 1.123(19) \\
\hline\hline
\end{tabular}
\label{tab:z_final}
\end{center}
\end{table}

The relations $Z_V=Z_A$ and $Z_S=Z_P$ for lattice chiral fermions are verified. 
These relations are better satisfied in the RI/SMOM scheme than in the RI/MOM scheme as shown in Fig.~\ref{fig:zv_massless}
and Fig.~\ref{fig:zsp}.
This is expected since the RI/SMOM scheme suppresses more nonperturbative effects by using symmetric momentum configurations.

For $Z_q$ and $Z_T$, the systematic uncertainties from their conversion ratios to the $\msbar$ scheme are larger 
if the RI/SMOM scheme is used instead of the RI/MOM scheme. In using both schemes for these two RCs, a renormalization window
can be found after the perturbative running to 2 GeV, 
in which a straightforward linear extrapolation in the initial scale $a^2p^2$ can be done. Our final results for matching
to the $\msbar$ scheme are
from the RI/MOM scheme since it gives more precise numbers.

For $Z_S$ and $Z_P$, the conversion ratio from the RI/SMOM scheme to the $\msbar$ scheme converges much faster than that from
the RI/MOM scheme to $\msbar$. 
However in using the RI/SMOM scheme we do not find a broad window in the initial momentum scale for $Z_S$ and $Z_P$
after the perturbative running,
in which both the nonperturbative effects and lattice discretization effects are small. That is to say, unlike for $Z_q$ and $Z_T$
we do not see a clear linear dependence on $a^2p^2$ in $Z_{S/P}^\msbar(2\mbox{ GeV}; a^2p^2)$ after running it from $a^2p^2$ to 2 GeV.
We tried several ansatzes Eqs.(\ref{eq:fx1})-(\ref{eq:fx5}) to model the behavior of $Z_{S/P}^\msbar(2\mbox{ GeV}; a^2p^2)$.
This leads to the large uncertainties
in our calculated $Z_S$ and $Z_P$ by using the RI/SMOM scheme. This reduced upper edge of the renormalization 
window may be from the usage of link smearing
as discussed in Ref.~\cite{Arthur:2013bqa}. We are interested in checking this on thin link configurations in the future.
  
In Refs.~\cite{Gorbahn:2010bf,Bell:2016nar}, two loop calculations of certain Green functions with bilinear quark operator insertions
were performed for a general momentum configuration parametrized 
by $\omega=q^2/p^2$. For example, $\omega=0$ and $1$ correspond to the exceptional and symmetric momentum configuration respectively.
The convergence in the perturbative series for the conversion to the $\msbar$ scheme for 
different operators behavior differently as a function of $\omega$.
Thus one may want to use a different $\omega$ for a different operator to mostly shrink the truncation error in the conversion ratio.
 
\section*{Acknowledgements}
We thank RBC-UKQCD collaborations for sharing the domain wall fermion configurations. 
This work is partially supported
by the National Science Foundation of China (NSFC) under Grants 11575197,
11575196, 11335001, 11405178, U1632104 and by the U.S.\ DOE Grant No.\ DE-SC0013065.
YC and ZL acknowledge the support of NSFC and DFG (CRC110).
This research used resources of the Oak Ridge Leadership Computing Facility at the Oak Ridge National Laboratory, 
which is supported by the Office of Science of the U.S. Department of Energy under Contract No. DE-AC05-00OR22725. 
This work used Stampede time under the Extreme Science and Engineering Discovery Environment (XSEDE)~\cite{XSEDE}, 
which is supported by National Science Foundation grant number ACI-1053575. We also thank National Energy Research 
Scientific Computing Center (NERSC) for providing HPC resources that have contributed to the research results reported within this paper. 
We acknowledge the facilities of the USQCD Collaboration used for this research in part, 
which are funded by the Office of Science of the U.S. Department of Energy.

\end{document}